\documentclass{emulateapj}
\usepackage{apjfonts}
\usepackage{graphicx}
\usepackage{amsmath}

\newcommand{\lsim}{\lower0.6ex\vbox{\hbox{$ \buildrel{\textstyle <}\over{\sim}\ $}}}
\newcommand{\gsim}{\lower0.6ex\vbox{\hbox{$ \buildrel{\textstyle >}\over{\sim}\ $}}}
\newcommand{\beq}{\begin{equation}}
\newcommand{\eeq}{\end{equation}}

\newcommand{\Omegam}{\Omega_{\mathrm{M}}}

\newcommand{\Omegade}{\Omega_{\mathrm{DE}}}

\newcommand{\omegam}{\omega_{\mathrm{M}}}
\newcommand{\omegab}{\omega_{\mathrm{B}}}

\newcommand{\dr}{\Delta_{\mathcal{R}}^{2}}
\newcommand{\wzero}{w_{0}}
\newcommand{\wa}{w_{\mathrm{a}}}
\newcommand{\apiv}{a_{\mathrm{p}}}
\newcommand{\wpiv}{w_{\mathrm{p}}}

\newcommand{\zpiv}{z_{\mathrm{piv}}}

\newcommand{\Da}{D_{\mathrm{A}}}
\newcommand{\thetas}{\theta_{\mathrm{s}}}

\newcommand{\zp}{z_{\mathrm{p}}}

\newcommand{\dd}{\mathrm{d}}

\newcommand{\Dl}{D_{\rm L}}

\newcommand{\dmlens}{\delta\mu_{\mathrm{lens}}}
\newcommand{\dmint}{\delta\mu_{\mathrm{int}}}

\newcommand{\siglens}{\sigma_{\mathrm{lens}}}
\newcommand{\sigint}{\sigma_{\mathrm{int}}}
\newcommand{\sigizero}{\sigma_{\mathrm{int},0}}

\newcommand{\sigmui}{\sigma_{\mu,\mathrm{i}}}
\newcommand{\meanmui}{\bar{\mu}_{\mathrm{i}}}

\newcommand{\nspec}{N_{\mathrm{spec}}}
\newcommand{\nspeci}{N_{\mathrm{spec}}^{\mathrm{i}}}
\newcommand{\nphoti}{N_{\mathrm{phot}}^{\mathrm{i}}}

\newcommand{\gi}{g_{\mathrm{i}}}
\newcommand{\fom}{\mathcal{A}}

\begin{document}
\submitted{The Astrophysical Journal, submitted}
\vspace{1mm}

\shortauthors{Zentner \& Bhattacharya}
\shorttitle{Supernova Ia Cosmology With Photometric Surveys}

\title{
Utilizing Type Ia Supernovae in a Large, Fast, Imaging Survey to Constrain Dark Energy
}

\author{
Andrew R. Zentner and Suman Bhattacharya
}
 
\vspace{2mm}


\begin{abstract}

We study the utility of a large sample of type Ia supernovae that might be observed 
in an imaging survey that rapidly scans a large fraction of the sky 
for constraining dark energy.  We consider both the information 
contained in the traditional luminosity distance test as well 
as the spread in Ia supernova fluxes at fixed redshift induced by 
gravitational lensing.  As would be required from an imaging survey, we include a treatment 
of photometric redshift uncertainties in our analysis.  Our primary result is 
that the information contained in the mean distance moduli of supernovae Ia 
and the dispersion of supernova Ia distance moduli 
complement each other, breaking a degeneracy between the present 
dark energy equation of state and its time variation without 
the need for a high-redshift ($z \gsim 0.8$) supernova sample.  
Including lensing information also allows for some internal 
calibration of photometric redshifts.  
To address photometric redshift uncertainties, we present dark energy constraints as 
a function of the size of an external set of spectroscopically-observed supernovae 
that may be used for redshift calibration, $\nspec$.    
Depending upon the details of potentially-available, external supernova data sets, we 
find that an imaging survey can constrain the dark energy equation of state at the 
epoch where it is best constrained $\wpiv$, with a 1$\sigma$ error of 
$\sigma(\wpiv) \approx 0.03-0.09$.  
In addition, the marginal improvement in the error $\sigma(\wpiv)$ from an increase 
in the spectroscopic calibration sample drops once 
$\nspec \sim \mathrm{a}\ \mathrm{few} \times 10^{3}$.  This result is important 
because it is of the order of the size of calibration samples likely to 
be compiled in the coming decade and because, for samples of this size, the 
spectroscopic and imaging surveys individually place comparable constraints 
on the dark energy equation of state.  In all cases, it is best to calibrate photometric 
redshifts with a set of spectroscopically-observed supernovae with relatively more objects 
at high redshift ($z \gsim 0.5$) than the parent sample of imaging supernovae.

\end{abstract}



\affiliation{Department of Physics \& Astronomy, The University of Pittsburgh, Pittsburgh, PA 15260}

\section{INTRODUCTION}
\label{section:introduction}

Type Ia Supernovae (SNeIa) can be used as well-calibrated standard candles with 
relatively little dispersion in their intrinsic luminosities 
\citep[e.g.,][]{phillips93,hamuy_etal95,riess_etal96,prieto_etal06,guy_etal07,jha_etal07}.  
This has allowed for recent measurements of the relation between 
cosmological distance and redshift, which provide 
the strongest contemporary evidence for an accelerating cosmological 
expansion \citep[see][for recent applications]{astier_etal06,riess_etal07,wood-vasey_etal07}.  
Several studies have examined the spread in observed fluxes (rather than their average) 
due to the magnification of SNeIa as a source of bias 
in cosmological parameter extraction from the distance--redshift test 
\citep{sasaki87,linder_etal88,kantowski_etal95,wambsganss_etal97,
holz98,wang99,barber00,holz_linder05}, 
a cross-check of lensing shear maps \citep[][and references therein]{cooray_etal06}, 
or as an interesting signal in its own right 
\citep{metcalf99,dodelson_vallinotto06,cooray_etal06,albrecht_etal06}.

The present paper is inspired by these studies and other efforts to explore how SNeIa might be 
utilized as part of a large, photometric survey 
\citep[e.g.,][]{albrecht_etal06,zhan_etal08,hannestad_etal08,zhang_chen08}.  
The aforementioned studies considered the lensing of a relatively small ($ \lsim 10^3$) 
spectroscopic sample of SNeIa, with redshifts that are known precisely 
\citep[with errors $\lsim 10^{-4}$, as prescribed by][]{huterer_etal04}.  
We examine the utility of a significantly larger ($\sim 10^6$) photometric sample of supernovae 
that may be collected by a future, wide, fast, imaging 
survey such as the proposed Large Synoptic Survey Telescope (LSST)\footnote{{\tt http://www.lsst.org}} 
in which the vast majority of SNeIa will not have accompanying spectroscopic observations.  
It is thought that the identification and determination of 
photometric redshifts for SNeIa can be done relatively reliably 
\citep{pinto_etal04,wang_etal07}.  Consequently, the dispersion 
in SNeIa fluxes at fixed {\em photometric} redshift has three sources:  
the intrinsic dispersion of SNeIa, including errors in calibrating the standard 
candles and applying K-corrections and extinction corrections; 
photometric redshift errors; and any dispersion induced by gravitational lensing.  
Moreover, each of these factors are thought to contribute to the total dispersion at 
fixed photometric redshift at comparable levels (see \S~\ref{section:methods}).  
The implication is that a very large sample of supernovae from an imaging survey 
can be used to overwhelm the intrinsic dispersion, thus 
enabling the measurement of SNeIa dispersion induced by photometric 
redshift errors and gravitational lensing which should shed light on both 
SNeIa photometric redshifts and cosmological parameters at relatively low redshift.

The modest goal of this paper is to make preliminary 
estimates of the utility of such a photometric SNeIa sample 
to learn about cosmology from the dispersion among SNeIa fluxes  
in conjunction with the traditional distance--redshift test.  
We find that the spread in observed SNeIa fluxes at fixed redshift 
provides constraints that are not, by themselves, competitive 
with other probes (such as weak lensing shear), 
but are yet interesting, subject to unique systematic errors, 
and obtained without significant additional 
observational effort if such a sample is to 
be used to perform the traditional distance--redshift 
test or to map baryon acoustic oscillations 
\citep[as studied in][]{albrecht_etal06,zhan_etal08}.  
In fact, because such a measurement will be most constraining on dark 
energy at low redshift ($z \sim 0.2-0.4$) it will have some degree of 
complementarity with other measures, such as cosmic shear which best 
constrains the dark energy equation of state near $z \sim 0.7$.  
More importantly, the information contained in the traditional distance--redshift 
test (most sensitive to the dark energy equation of state near $z \sim 0.1$) and 
the dispersion of SNeIa brightnesses complement {\em each other} and help to make 
cosmology with photometric SNeIa more robust to uncertainties in photometric redshifts.

Of course, the lack of spectroscopy is a {\em major} complication in using SNeIa 
from a photometric survey.  As the use of photometric SNeIa 
samples for cosmology requires some treatment of 
photometric redshift uncertainties, we also address 
the dependence of cosmological constraints upon the 
calibration of SNeIa photometric redshifts.  In particular, we find that utilizing the 
dispersion in SNeIa distance moduli allows for mild self-calibration in the 
uncertainties of photometric redshift distributions.   Perhaps more interestingly, 
we present results for dark energy constraints as a function of the size of a 
spectroscopic sample of SNeIa that can be used to calibrate photometric redshifts for 
two different assumptions about the redshift distribution of this spectroscopic  
calibration set.  This is a convenient way to express cosmological constraints as 
a function of prior knowledge of the photometric redshift distribution of SNeIa.  
Interestingly, we find that if the goal is to seek the best constraint on the 
equation of state parameter at the epoch where it is best constrained by the data 
(the so-called ``pivot'' equation of state, see \S~\ref{section:methods}, 
results are similar for a constant equation of state model), then there 
is decreasing marginal value in expanding the size of the spectroscopic 
calibration set beyond a few thousand SNeIa, a number of SNeIa that should be achievable 
in the coming decade.  However, this statement does depend upon model details and 
upon the assumption that it is the pivot equation of state that is of most interest.

There are several important caveat to our analysis.  We neglect the, 
perhaps considerable, additional complications in supernova type identification 
and calibration arising from the lack of spectroscopy.  Aside from 
neglect of several potentially-important observational realities, our calculation 
is also an idealized one.  We have treated the influence of gravitational lensing on 
observed fluxes in a simplified manner, in large part because current analytic models 
for the proposed lensing signal are not adequate for future surveys and a full treatment 
of this effect is computationally demanding.  However, 
part of the purpose of this paper is to demonstrate that such a signal does contain 
valuable information and to emphasize that as we yet lack the theoretical tools 
needed to perform a robust analysis of future data, developing such 
tools should be a high priority in the run-up to forthcoming photometric surveys.

The present paper is organized as follows.  In \S~\ref{section:methods} 
we describe the our treatment of SNeIa observables and photometric redshifts 
as well as our methods for forecasting cosmological 
parameter constraints.  We present our primary results in 
\S~\ref{section:results}.  We discuss our results, including 
implications and important caveats in \S~\ref{section:discussion}.

\section{Basis and Methods}
\label{section:methods}

The average distance moduli $\mu$, of high-redshift supernovae as a function of 
redshift provide the most direct evidence for cosmological acceleration.  In the 
absence of any lensing, the distance modulus of a standard candle 
at true redshift $z$ is $\mu = 5 \log [\Dl(z)/10~\mathrm{pc}]$, where 
$\Dl(z)$ is the luminosity distance to redshift $z$.  We explore both the mean and 
dispersion in distance moduli of SNeIa measured in a future, large photometric survey.  
We assume a survey that covers $ \approx 20,0000$~deg$^2$ 
to a redshift of $z \sim 0.8$ for $10$ years as might be achieved 
with the LSST instrument and survey strategy.  We take 
a SNeIa redshift distribution $\dd n/\dd z$ as in \citet{zhan_etal08}, which 
rises to $z \sim 0.5$ ($ \dd n/\dd z \propto z^{2.1}$ for $z \ll 0.5$) 
and declines rapidly thereafter ($\dd n/\dd z \propto \exp[-32(z-0.5)^2]$ for $z \gg 0.5$).  
The rate at which supernovae may be discovered and 
the efficiency with which they will produce light curves suitable for 
calibration and reliable photometric redshift determination is uncertain and depends upon 
the exposure strategy of any such survey.  For simplicity, we 
choose a fiducial survey containing one million SNeIa.  We note in passing that this 
normalization is significantly more ambitious than the $3 \times 10^5$ SNeIa assumed for the 
hypothetical, stage IV, ground-based survey explored by the Dark Energy Task Force 
\citep[DETF,][]{albrecht_etal06}, but significantly less ambitious than the 
$7.4$~million SNeIa assumed in the exploration of SNeIa as a tracer of 
baryon acoustic oscillations by \citet{zhan_etal08}.  In the absence of additional 
prior constraints on cosmological parameters and external calibration samples of SNeIa (see details below) 
parameter uncertainties would scale with sample size as $\sigma \propto 1/\sqrt{N_{\mathrm{SNe}}}$; 
however, our constraints scale more slowly because we assume priors on cosmological parameters 
as well as the availability of additional spectroscopic SNeIa samples that improve the 
calibration of SNeIa properties and redshifts.

\begin{figure*}[t!]
\includegraphics[width=16.5cm]{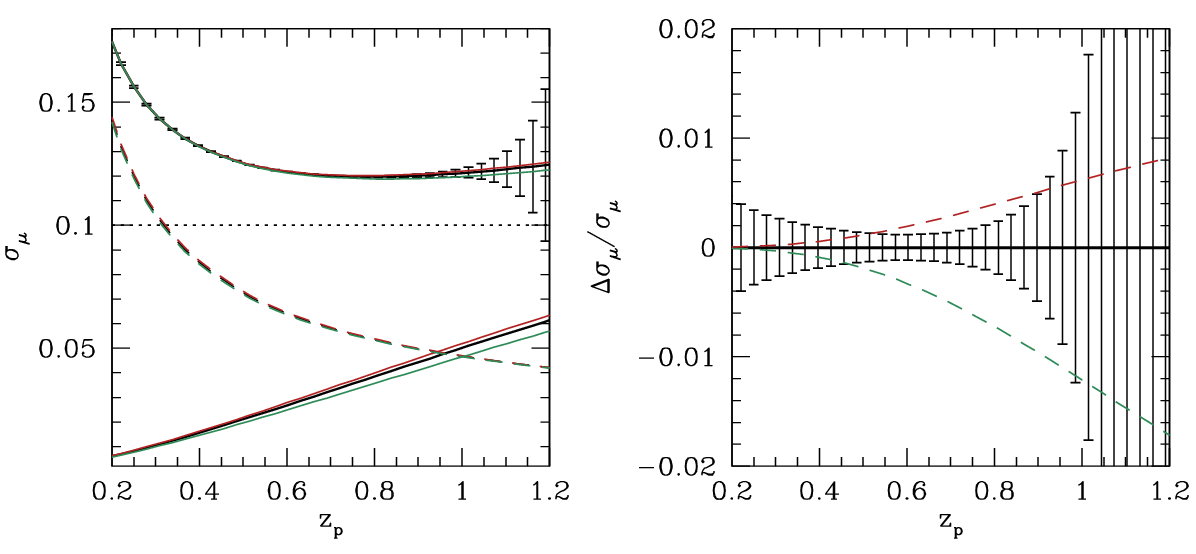}
\caption{
The dispersion in supernova distance moduli.  
{\it Left:}  Dispersion as a function of photometric redshift for 
SNeIa in 51 bins of photometric redshift.  
The {\em horizontal, dotted} line represents the assumed intrinsic dispersion 
level of $\sigint=0.1$.  A set of three, nearly indistinguishable {\em dashed} lines 
decrease as a function of redshift.  These lines represent the contribution to the 
dispersion from the first two terms in Eq.~(\ref{eq:disp}).  The central 
line corresponds to our fiducial cosmological model while the upper (lower) 
line corresponds to the same cosmology but with $\wzero=-1.1$ ($\wzero=-0.9$).  
A set of three {\em solid} lines increases monotonically with redshift.  These 
represent the contribution to the net dispersion from gravitational lensing.  
The central line is for our fiducial cosmology, while the upper (lower) line 
has $\wzero=-1.1$ ($\wzero=-0.9$).  The {\em uppermost, solid} lines 
represent the total dispersion in Eq.~(\ref{eq:disp}).  The central, upper, 
and lower lines correspond to models with $\wzero=-1$, $\wzero=-1.1$, and 
$\wzero=-0.9$ respectively, while the errorbars correspond to the statistical 
error on a determination of each of the 51 $\sigmui$ from a sample of one million SNeIa.  
{\it Right:}  Residuals of the dispersion in distance modulus with respect to 
our fiducial cosmological model with $\wzero=-1$.  The upper dashed line is 
the dispersion in a cosmology with $\wzero=-1.1$ and the lower line is the 
dispersion with $\wzero=-0.9$.  The errorbars are the errors on the measurement 
of the dispersion in each of 51 photometric redshift bins as in the left panel.  
This figure is modeled after Figure~1 of \citet{dodelson_vallinotto06} and demonstrates that 
the dispersion in standard candle distance moduli that may be measured in a forthcoming, 
large, photometric survey like the LSST will undertake, has some leverage to distinguish 
dark energy equations of state.
}
\label{fig:sigmu}
\end{figure*}

Each observed supernova will be assigned a photometric redshift $\zp$ that is significantly 
less reliable than could be achieved with spectroscopy.  We model photometric 
redshift uncertainties as follows \citep[][]{ma_etal06}.  We assume a distribution of 
photometric redshifts given a spectroscopic redshift $P(\zp|z)$, that is a Gaussian with an 
offset from the true redshift $b_z(z)$ and dispersion $\sigma_z(z)$, each of which 
may depend upon redshift.  Recent studies indicate that $b_z \approx 0$ and 
$\sigma_z \approx \sigma_{z0}(1+z)$ with $\sigma_{z0} \approx 0.01$ should be 
achievable \citep{pinto_etal04,albrecht_etal06,wang_etal07}.  In practice, it is necessary to 
allow $b_z$ and $\sigma_z$ to have significant freedom to vary as a function of 
$z$.  We do this by representing the functions $b_z$ and $\sigma_z$ by the values of 
these functions tabulated at $n_{\mathrm{pz}}$ redshifts spaced evenly in $z$ from 
$z=0$ to $z=1.5$ and evaluating these functions at points between the redshifts at 
which they are tabulated using linear interpolation.  
Thus there are $2n_{\mathrm{pz}}$ photometric redshift parameters.  
We set $n_{\mathrm{pz}}$ by requiring that $n_{\mathrm{pz}}$ be large enough that 
when all $2n_{\mathrm{pz}}$ parameters are allowed to vary without prior knowledge, 
the constraining power of the data converges to a minimum value 
corresponding to complete lack of redshift information.  
In practice, our results converge to this minimum with $n_{\mathrm{pz}} \approx 15$ and to 
be conservative, we fix $n_{\mathrm{pz}}=31$ as in the lensing study of \citet{ma_etal06} 
for the remainder of this paper.

Of course, there will be {\em some} prior knowledge of $P(\zp|z)$ and in principle each of 
the photometric redshift model parameters will have its own prior.  In fact, some calibration 
of the photometric redshift distribution is needed in order to utilize SNeIa from an 
imaging survey as we demonstrate explicitly in the following section.  For simplicity, 
we quantify the influence of priors using two one-parameter models that are reasonable, 
but not exhaustive of all possibilities.  We assume a sample of $\nspec$ SNeIa with 
known spectroscopic redshifts.  We further assume that this calibration sample either 
follows the overall distribution $\dd n/\dd z$, or is spaced uniformly in redshift 
from $z=0.1$ to $z=1.7$ as might be expected from a Joint Dark Energy Mission (JDEM) SNeIa 
sample\footnote{The Supernova Acceleration Probe (SNAP) is an example, see {\tt http://snap.lbl.gov}} 
that might be a contemporary of a large-scale photometric survey \citep{kim_etal04,albrecht_etal06}.  
We also incorporate the influence of a nearby sample of SNeIa with spectroscopic redshifts that 
help to serve as a low-redshift anchor for the distance--redshift test.  For this, we take the 
{\em nearby} SNeIa sample to contain 500 objects uniformly distributed in the redshift interval 
$z=0.03-0.08$ with redshift errors of $\sigma_z=10^{-4}$ as expected from the ongoing Nearby Supernova Factory 
\citep[][]{aldering_etal02,copin_etal06,albrecht_etal06}\footnote{{\tt http://snfactory.lbl.gov}} 
experiment.

Any calibration set would determine the bias, $b_z^{\mathrm{i}}$, and dispersion, 
$\sigma_{\mathrm{z}}^{\mathrm{i}}$, of the photometric redshift distribution in redshift 
bin $\mathrm{i}$ with uncertainties of 
$\sigma_{b_z^{\mathrm{i}}}=\sigma_{\mathrm{z}}^{\mathrm{i}}/\sqrt{\nspeci}$ 
and $\sigma_{\sigma_{\mathrm{z}}^{\mathrm{i}}}=\sigma_{\mathrm{z}}^{\mathrm{i}}/\sqrt{2\nspeci}$ 
where $\nspeci$ is the number of spectroscopic SNeIa that fall in bin i.  
We include the influence of prior knowledge of the photometric redshift distribution 
using such priors.  We assume a fiducial model set by $b_z=0$ and $\sigma_z=\sigma_{z0}(1+z)$ 
with $\sigma_{z0}=0.01$.  We reiterate that, unless otherwise stated, 
we allow for uncertainty in the photometric redshift distribution as a 
function of redshift through the $2n_{\mathrm{pz}}$ photometric redshift parameters and {\em do not} 
assume that the photometric redshift parameters are fixed at the fiducial values 
(in contrast to, for example, \citealt{hannestad_etal08}, who assume perfect knowledge of SNeIa redshifts and 
\citealt{zhan_etal08} who do not consider calibration of the bias in photometric redshifts).

For computational convenience, we compute the observable properties of supernovae in 
bins of photometric redshift.  Given the distribution $P(\zp|z)$, the true 
redshift distribution of SNeIa in photometric redshift bin i is
%
%
\beq
\label{eq:dnidz}
\frac{\dd n_{\mathrm{i}}}{\dd z} = \int_{\zp^{\mathrm{low}}}^{\zp^{\mathrm{high}}} 
\frac{\dd n}{\dd z} P(\zp|z) \dd \zp,
\eeq
where $\zp^{\mathrm{low}}$ and $\zp^{\mathrm{high}}$ are the lower and upper boundaries of photometric 
redshift bin i respectively.  The total number of SNeIa in photometric redshift bin i is 
$\nphoti = \int (\dd n_{\mathrm{i}}/\dd z) \dd z$ and we define the normalized redshift 
distribution of SNeIa in bin i as $\gi(z) = (\dd n_{\mathrm{i}}/\dd z)/\nphoti$.  In practice, 
we take 51 bins spaced equally in redshift from $z=0$ to $z=1.5$.  Though convenient 
considering observables averaged within redshift bins reduces information content relative to the total 
information available.  We have explored various binning schemes and find that finer binning 
is unnecessary as our results are already converged with this number of bins.  
In particular, increasing the number of bins by a factor of $3$ gives only $\sim 4\%$ 
better constraints on the pivot dark energy equation of state parameter defined below.

Consider an effective distance modulus defined for convenience to be the 
difference between apparent magnitude and an assumed average absolute magnitude.  
Any observed supernova has an effective distance modulus 
%
%
\beq
\label{eq:dmod}
\mu = \Delta + 5 \log \Bigg(\frac{\Dl}{10~\mathrm{pc}}\Bigg) + \dmint + \dmlens, 
\eeq
where $\dmint$ is the variation in apparent magnitude 
due to the intrinsic variations in SNeIa absolute magnitudes, 
$\dmlens$ is the contribution from gravitational lensing, 
and $\Delta$ is some constant offset that is applied to all supernovae and 
represents a deviation from the assumed average absolute magnitude which includes 
errors in $H_0$.  In what follows, we set $\Delta$ to vary freely.  
The average distance modulus of SNeIa in the i$^{\mathrm{th}}$ photometric redshift bin is 
%
%
\beq
\label{eq:meanmu}
\meanmui = \Delta + 5 \int \dd z \ \gi(z) \log \Bigg[ \frac{\Dl(z)}{10~\mathrm{pc}} \Bigg]
\eeq
and the standard distance--redshift test is an application of this 
relation.   In addition to distance modulus, 
the Greek letter ``$\mu$'' is also the conventional symbol for the 
lensing magnification.  In order to avoid confusion, we will {\em not} refer to the 
lensing magnification directly but only to the shift in distance modulus induced by 
weak gravitational lensing.

The dispersion in the effective distance modulus $\sigmui$, is given by
%
%
\begin{eqnarray}
\label{eq:disp}
\sigmui^2 & = & 25\Bigg(\int \dd z\ \gi(z) \log^2 \Bigg(\frac{\Dl}{10~\mathrm{pc}} \Bigg) 
- \Bigg[ \int \dd z\ \gi(z) \log \Bigg(\frac{\Dl}{10~\mathrm{pc}}\Bigg) \Bigg]^2 \Bigg) \nonumber \\
          &   & + \int \dd z\ \gi(z) \sigint^2(z) + \int \dd z\ \gi(z) \siglens^2(z),
\end{eqnarray}
where $\sigint$ is the intrinsic dispersion of SNeIa brightnesses, 
which may itself vary with redshift, and $\siglens$ is the dispersion 
induced by gravitational lensing.  The first two terms in parenthesis in Eq.~(\ref{eq:disp}) 
represent the spread in observed SNeIa distance moduli due to the spread in photometric 
redshifts.

In the weak lensing limit (convergence $\kappa \ll 1$), 
the dispersion due to lensing is \citep{bernardeau_etal97,valageas00,dodelson_vallinotto06}
%
%
\begin{eqnarray}
\label{eq:lens}
\siglens^2(z) &= & \frac{225 \Omegam^2H_0^2}{8\pi \mathrm{ln}^2(10)}\int_0^z \dd z' \frac{W^2(z',z)}{H(z)} \nonumber \\
              & & \times \int \dd k \ k P(k,z') \ {\mathcal{W}}_{\mathrm{TH}}^2(\Da k \thetas),
\end{eqnarray}
where $W(z',z)=H_0 \Da(z') \Da(z',z)/\Da(z)$, $\Da(z)$ is the angular diameter distance to redshift $z$, and 
$\Da(z',z)$ is the angular diameter distance between redshifts $z'$ and $z$.  The quantity 
$P(k,z)$ is the matter power spectrum and we evaluate it using the relation of \citet{smith_etal03}.  
The function $\mathcal{W}_{\mathrm{TH}}(x) = 2J_1(x)/x$ in Eq.~(\ref{eq:lens}) 
is a smoothing function that arises by considering the 
magnification averaged over an angular tophat of radius $\thetas \ll 1$.  
In practice, we set $\thetas=1''$, but the choice is relatively unimportant 
in that $\Da k \thetas \ll 1$ and so $\mathcal{W}_{\mathrm{TH}}(\Da k \thetas) \simeq 1$ 
for relevant choices of $\thetas$.  
We assume the lensing to be uncorrelated \citep{metcalf99,cooray_etal06,dodelson_vallinotto06} and 
take the $\sigmui$ as an independent set of observables that can be extracted from the 
observed SNeIa population.

Figure~\ref{fig:sigmu} shows the dispersion in distance modulus in 51 photometric redshift 
bins as a function of the center of the photometric redshift bin $\zp$ in our fiducial 
cosmological model.  Our fiducial model has $\omegam=0.13$, $\omegab=0.0223$, $\Omega_{k}=0$, 
$\dr=2.47 \times 10^{-9}$, and $n_s=1$.  We describe dark energy by its present energy 
density $\Omegade=0.76$ and a time-varying equation of state parameter $w(a)=\wzero+\wa(1-a)$, 
with $\wzero=-1$ and $\wa=0$ in our fiducial model.  For the purpose of this illustration, we take 
$\sigint=0.1$.  We will return to the intrinsic dispersion later.  Notice that the contributions to the 
dispersion due to the photometric redshift distribution and gravitational lensing are comparable and 
give the total dispersion a characteristic dependence upon redshift.  In particular, the dispersion decreases 
at low redshift and increases at high redshift due to the increased dispersion due to lensing.  
This makes it possible to use the dispersion to extract cosmological information without a 
good constraint on the level of the intrinsic dispersion.

We estimate constraining power using a Fisher matrix analysis.  
The Fisher matrix formalism is ubiquitous in cosmological parameter forecasting 
\citep[useful references include~][~the last three of which explore cases similar to the present paper]
{jungman_etal96,tegmark_etal97,seljak97,metcalf99,albrecht_etal06,zhan_etal08} 
and so we will not review the formalism here.  We take as our 
observables the set $\meanmui$ and $\sigmui$ with errors of 
$\sigma(\meanmui) = \sigmui/\sqrt{\nphoti}$ and 
$\sigma(\sigmui) = \sigmui/\sqrt{2\nphoti}$ respectively. 
We refer to constraints using the 
$\meanmui$ alone as ``luminosity distance test,'' and 
constraints from the $\sigmui$ alone as the ``dispersion test.''  
The Fisher matrix $F_{mn}$, approximates the covariance in model parameters 
locally about the maximum likelihood, so the indices run 
over the parameters in the parameter space we seek to constrain.  
The $1\sigma$ constraint on parameter $p$ is approximated by the square root 
of the diagonal component of the inverse Fisher matrix corresponding to this parameter.  
We denote this $\sigma(p) \simeq \sqrt{[F^{-1}]_{pp}}$.  Along with the parameters $\Delta$ 
and $\sigint$ describing SNeIa and the parameters describing the photometric redshift distribution, 
we vary the eight cosmological parameters in the previous paragraph about 
their quoted fiducial values.  We take independent, Gaussian priors on these parameters of 
$\sigma(\omegam)=0.004$, $\sigma(\omegab)=6 \times 10^{-4}$, $\sigma(\Omega_{k})=10^{-3}$, 
$\sigma(\ln \dr)=0.04$, and $\sigma(n_s)=0.02$, all of which are comparable to 
contemporary constraints on these parameters \citep[][]{komatsu_etal08}.

Absent significant direction from theory, it is not clear how to assess the constraining 
power of a dark energy program.  We use two reasonable metrics.  First, we consider 
the area of the marginalized 95\% error ellipse in the $\wzero$-$\wa$ plane $\fom$, 
as suggested by the DETF \citep[][the DETF actually uses $1/\fom$, see also 
\citealt{huterer_turner01} for a similar suggestion]{albrecht_etal06}.  Alternatively, 
it may be supposed that the goal of a dark energy experiment should be primarily to limit any 
deviation from a vacuum energy or cosmological constant, both of which have constant $w=-1$.  
In this case, it is interesting to examine the error on the equation of state parameter 
$w(a)$ at the epoch where it is most well constrained, the so-called pivot scale factor $\apiv$ 
or pivot redshift $\zpiv=1/\apiv-1$ \citep[e.g.,][]{huterer_turner01,hu_jain04,albrecht_etal06}.  
The pivot scale factor is $\apiv = 1 + [F^{-1}]_{\wzero \wa}/[F^{-1}]_{\wa \wa}$.  
The corresponding pivot equation of state is $\wpiv = \wzero + (1-\apiv)\wa$, so the 
error on $\wpiv$ is $\sigma^2(\wpiv) = [F^{-1}]_{\wzero \wzero} - [F^{-1}]_{\wzero wa}^2/[F^{-1}]_{\wa wa}$.  
The parameters $\wpiv$ and $\wa$ are uncorrelated and the transformation from 
a $\wzero$-$\wa$ parameterization preserves the area of the error ellipse in the two parameters, 
so $\fom \simeq  6.17  \pi \sigma( \wpiv ) \sigma( \wa )$.

\begin{figure*}[t!]
\includegraphics[width=8.2cm]{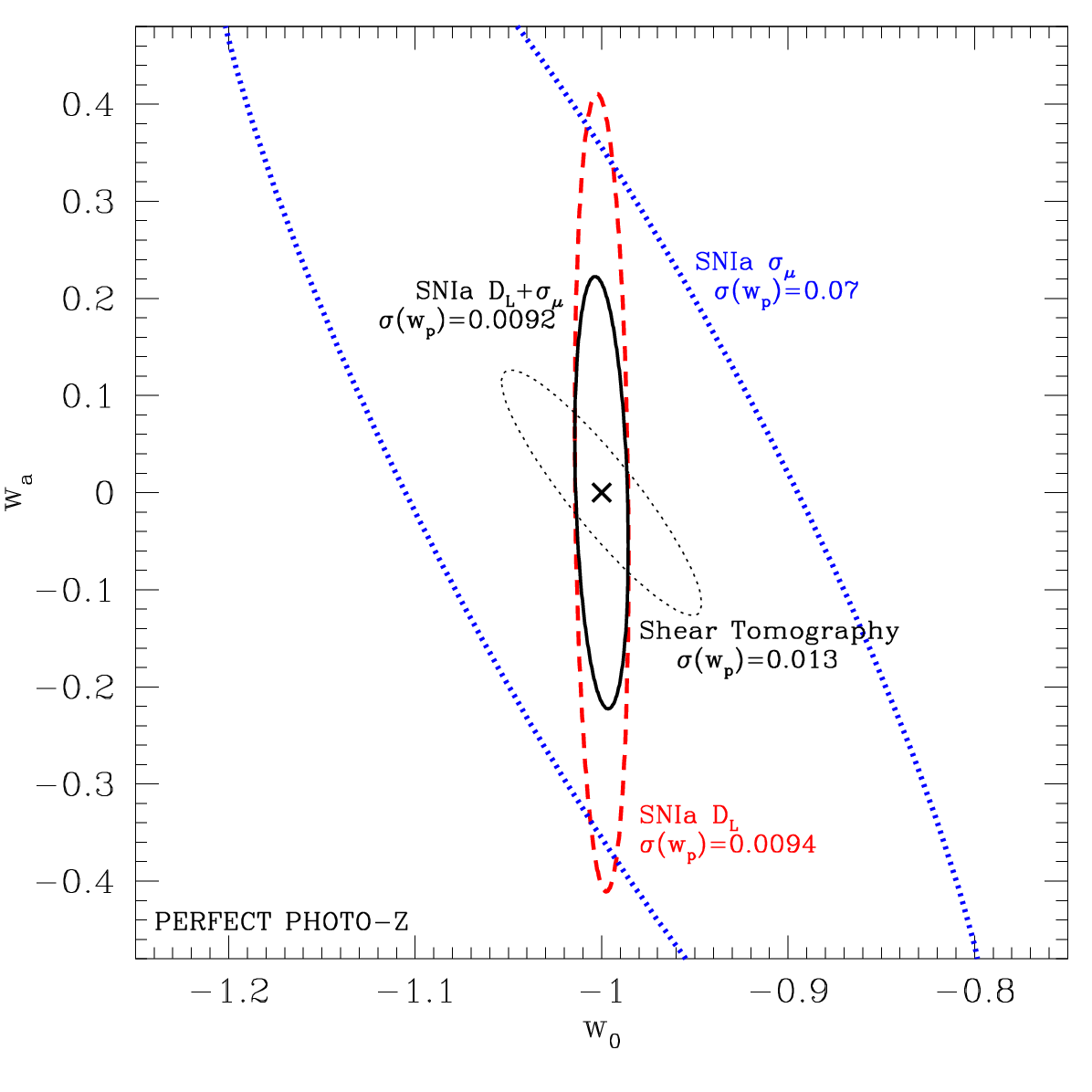}
\includegraphics[width=8.2cm]{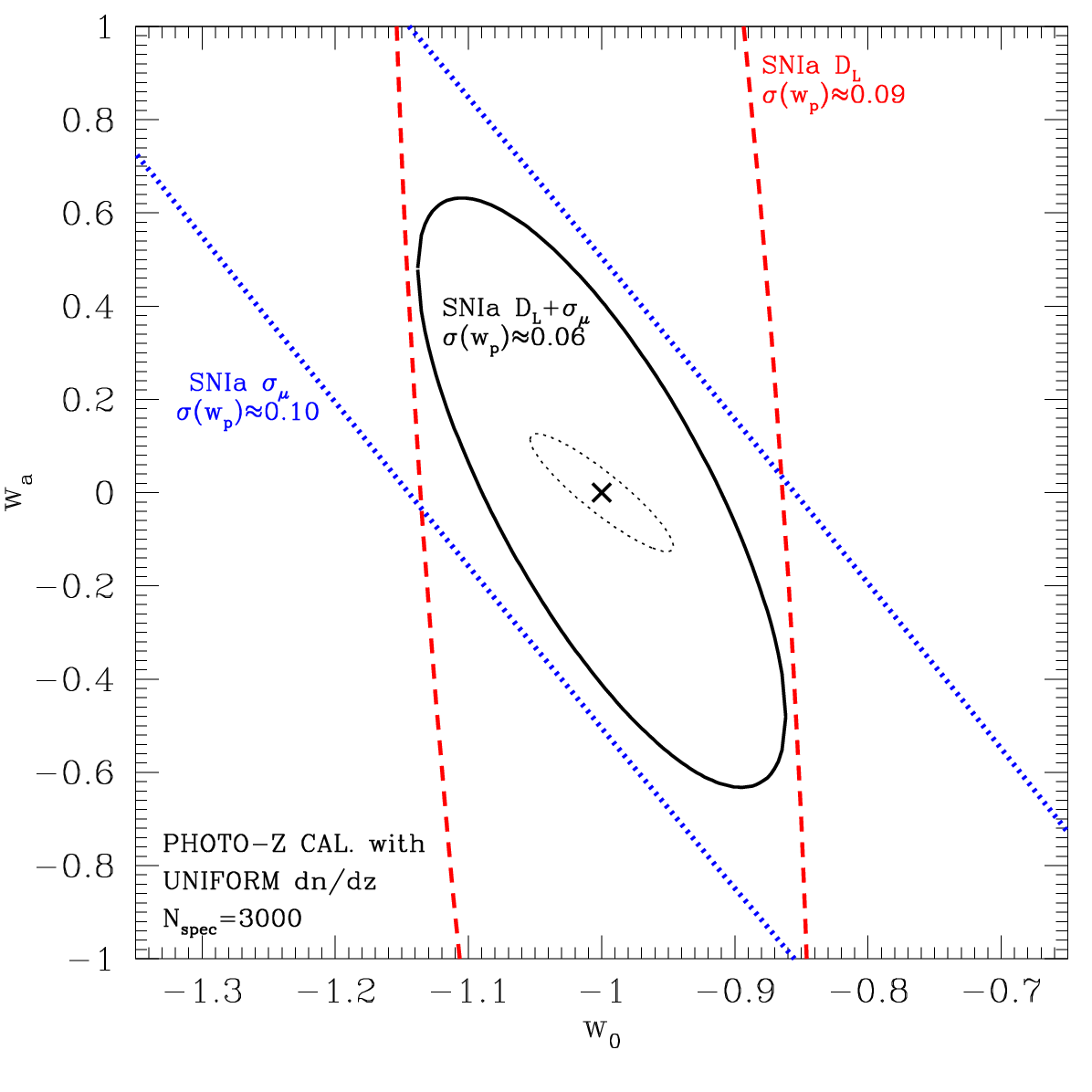}
\caption{
Constraint contours in the $\wzero$-$\wa$ plane.  The {\em left} panel shows 
1$\sigma$ constraint contours in the limit of perfect knowledge of the parameters of the 
photometric redshift model.  The {\em right} panel shows parameter constraints 
assuming priors on the photometric redshift parameters that would be obtained 
after calibrating to a spectroscopic sample of $\nspec=3000$ SNeIa distributed 
uniformly in redshift over the interval $0.1 \le z \le 1.7$.  
Note that the axes in the two panels span different ranges.  
In each panel, the cross marks the fiducial model with $\wzero=-1$ and $\wa=0$.  
The {\em outer, doshed} contours represent the dark energy constraints from the 
standard luminosity distance--redshift test [Eq.~(\ref{eq:meanmu})] alone.  
The {\em outer, dotted} contours represent constraints from the dispersion among 
SNeIa alone.  The {\em solid} contours show the combined constraints from 
both the mean and dispersion among SNeIa distance moduli.  
To set these constraints in context, the {\em inner, thin, dotted} contours 
in each panel show the constraints expected from galaxy weak lensing 
shear tomography in an LSST-like survey computed using the same fiducial 
model as described in \citet{zentner_etal08}.
}
\label{fig:w0wa}
\end{figure*}

Our estimate of parameter uncertainties stands as one 
of many abuses of the Fisher matrix that appear in the 
literature and this warrants some discussion.  
For one thing, we expect the likelihood to be non-Gaussian for an interesting range of 
parameter values \citep[though][also works in the Gaussian approximation, see also \S~\ref{section:discussion}]{metcalf99}.  
More importantly, we consider only the mean [Eq.~(\ref{eq:meanmu})] and 
dispersion [Eq.~(\ref{eq:disp})], rather than the full $\mu$ distribution.  
In principle, one would like to, and should, 
account for the full shape of the distribution of $\mu$, which is 
non-Gaussian in large part because the distribution of magnifications due to 
lensing is strongly non-Gaussian 
\citep[e.g.,][]{schneider_wagoner87,sasaki87,wambsganss_etal97,holz98,wang99,valageas00,munshi_jain00,wang_etal02}.  
One way to see that lensing should be non-Gaussian is to recognize that a minimum amount of 
de-magnification (or dimming) will occur when the null geodesic passes through an empty beam 
with density $\rho=0$.  The minimum magnification of a source at redshift $z_s$, 
corresponds to a maximum shift in distance modulus of 
$\delta \mu_{\mathrm{max}} = (15/2) \Omegam H_0 \int_0^{z_s} \dd z \  W(z,z_s)/H(z)$.  There is no 
corresponding upper limit to magnification or corresponding lower limit to $\dmlens$.

We use the Fisher matrix formalism and consider only the dispersion in $\mu$ as a matter of pragmatism.  
First, computing the distribution of magnifications due to lensing is computationally 
intensive and still subject to uncertainties in numerical modeling of nonlinear structure 
formation \citep[e.g.,][]{huterer_takada05,hagan_etal05,rudd_etal08} at levels that are important 
for forthcoming data.  Neglecting systematic issues, this problem could be circumvented if 
a reliable, analytic fitting form for the magnification distribution could be used.  
\citet{wang_etal02} provide such a fit, but we find that the \citet{wang_etal02} relation is 
neither sufficiently accurate to address forthcoming large data sets 
($\sim$~percent-level predictions are necessary), nor is it internally 
self-consistent below redshifts $z \sim 0.6$, where the probability density of 
magnification exhibits discontinuities and violates flux 
conservation\footnote{The redshifts at which these issues of inconsistency become important 
vary considerably with cosmological parameters.  These issues are alluded to in the work 
of \citet{dodelson_vallinotto06}, but they do not specify the shortcomings of the 
\citet{wang_etal02} fitting form.  The shortcomings are understandable in the sense that 
the present application was likely not foreseen by \citet{wang_etal02}, so it seems reasonable 
that these authors would not expend significant effort to calibrate their fit at low redshifts.}.  
This is particularly important for the present study because 
the bulk of SNeIa in any ground-based, photometric survey of a large fraction of the 
sky will be at redshifts $\lsim 0.6$.

If such a fitting form were available, the present study with $\sim 10^6$ SNeIa and 
$\sim 70$ parameters would still be computationally demanding.  
What one would like to do is to perform a Monte-Carlo analysis 
\citep[as was done in][]{dodelson_vallinotto06} in which random realizations of SNeIa fluxes 
are generated and to map out the likelihood during a random walk through the parameter 
space \citep[e.g.,][]{christensen_meyer01,christensen_etal01,knox_etal01,kosowsky_etal02}.  
In doing so, one would also treat the full lensing distribution, rather than taking the 
weak lensing limit (where convergence $\kappa \ll 1$ or $\dmlens \ll 1$), which does not 
account properly for the objects in the high-magnification tail of the lensing distribution.  
\citet{dodelson_vallinotto06} and \citet{sarkar_etal08a} have shown that working in this 
limit does induce biases in cosmological parameters inferred from any observational 
data sets, but does not significantly alter parameter constraints.  
Indeed, we note here that we have successfully repeated the 
Monte Carlo analysis of \citet{dodelson_vallinotto06}, finding similar results.  
We have also performed the same Monte Carlo analysis for our photometric survey in 
cases of a reduced parameter space where the photometric redshift distribution is 
assumed to be known perfectly and in which only $\Omegam$ and $\wzero$ vary.  In both cases, 
a Fisher matrix analysis yields constraints that are in 
rough agreement ($\sim 30\%$) with the more complete approach.
Detailed and precise theoretical predictions and detailed survey strategies 
and systematics analyses are not yet available, so we do not consider a 
calculation that incorporates a full Monte Carlo analysis of the lensing 
distribution and the parameter space to be warranted at present and, for practical 
reasons, we have not performed such a calculation.  However, any analysis of 
an observational data set must undertake such a calculation.

Rather than treating such a specific calculation, 
our point is to indicate that the dispersion among supernovae in any large photometric 
survey that also aims at using the SNeIa sample to perform the 
distance--redshift test, measure baryon acoustic oscillations, or even 
assess SNeIa systematics will be a source of meaningful information.  In fact, 
part of our point is that such information may be present in future 
data, but the theoretical tools to do a rigorous analysis of such data 
are lacking and must be developed.  
We return to many of these issues in \S~\ref{section:discussion}.

\section{Results}
\label{section:results}

This section contains our results regarding the utility of a large, photometric 
sample of SNeIa to constrain cosmological parameters.  Our primary interest is 
in the ability of SNeIa to constrain the dark energy equation of state, 
and so we focus our attention on $\wzero$, $\wa$, the pivot value of the 
equation of state parameter $\wpiv$, and the 95\% $\wzero$-$\wa$ ellipse area $\fom$.  
We illustrate the general utility of SNeIa from a large photometric survey, and in particular 
the utility of examining both the mean and dispersion among distance moduli, in 
Figure~\ref{fig:w0wa}.

The left panel of Fig.~\ref{fig:w0wa} shows constraints in the 
unrealistically-optimistic situation where 
the parameters of the photometric redshift model are known 
perfectly.  In reality, this cannot be the case, but this represents the limit of 
the best possible cosmological constraints that could be achieved with a photometric 
survey of supernovae.  We present results for a fiducial case with $10^6$ 
supernovae.  We reiterate that for a fixed redshift distribution and in the 
absence of external priors, constraints would scale with the total number of supernovae as 
$\sigma \propto 1/\sqrt{N_{\mathrm{SNe}}}$ and this could be used to scale 
constraints to approximate those from smaller or larger samples with similar 
redshift distributions \citep[e.g.,][]{albrecht_etal06,hannestad_etal08,zhan_etal08}.  
Our constraints scale somewhat more slowly due to the influence of the external 
priors we assume.

\begin{figure*}[t!]
\includegraphics[width=8.0cm]{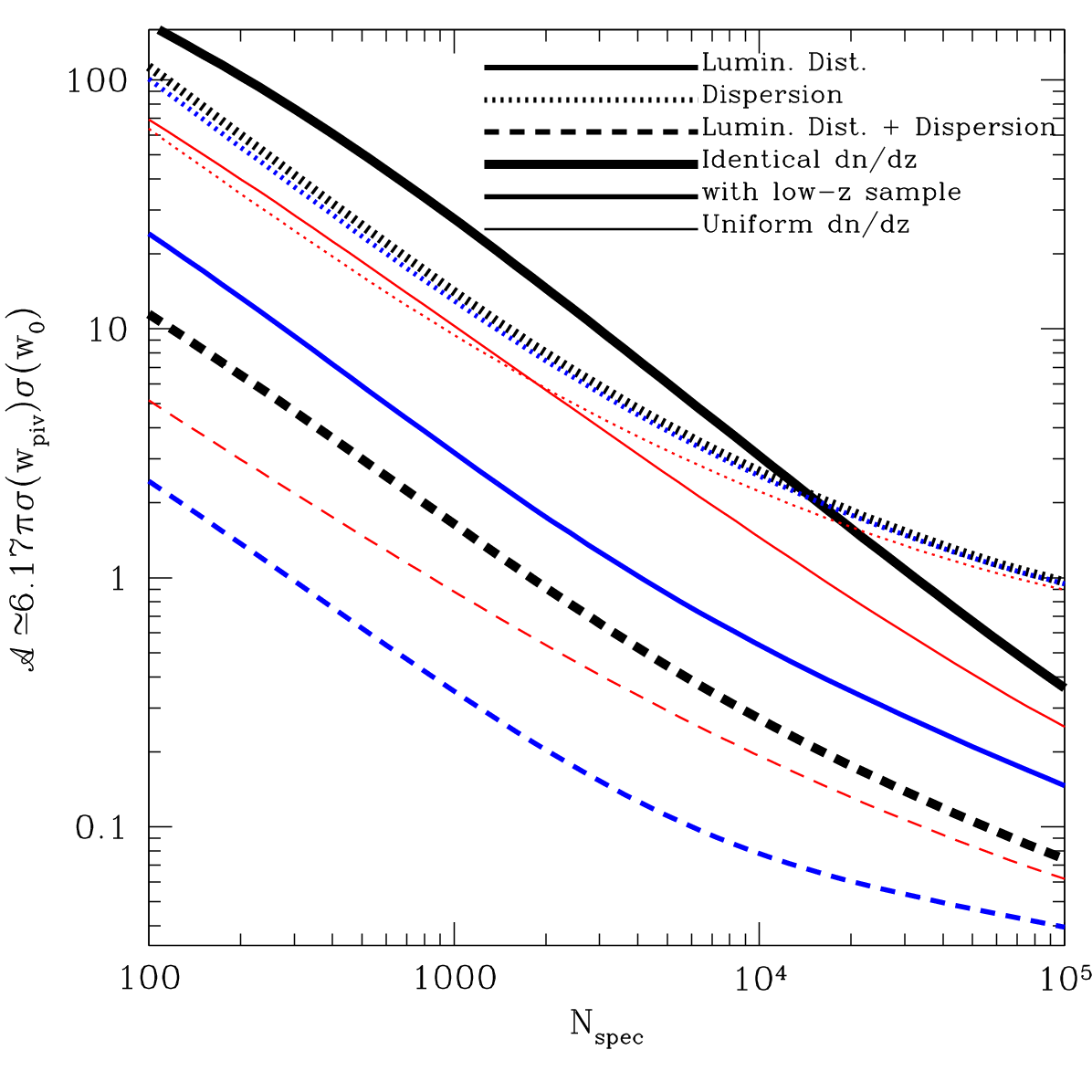}
\includegraphics[width=8.0cm]{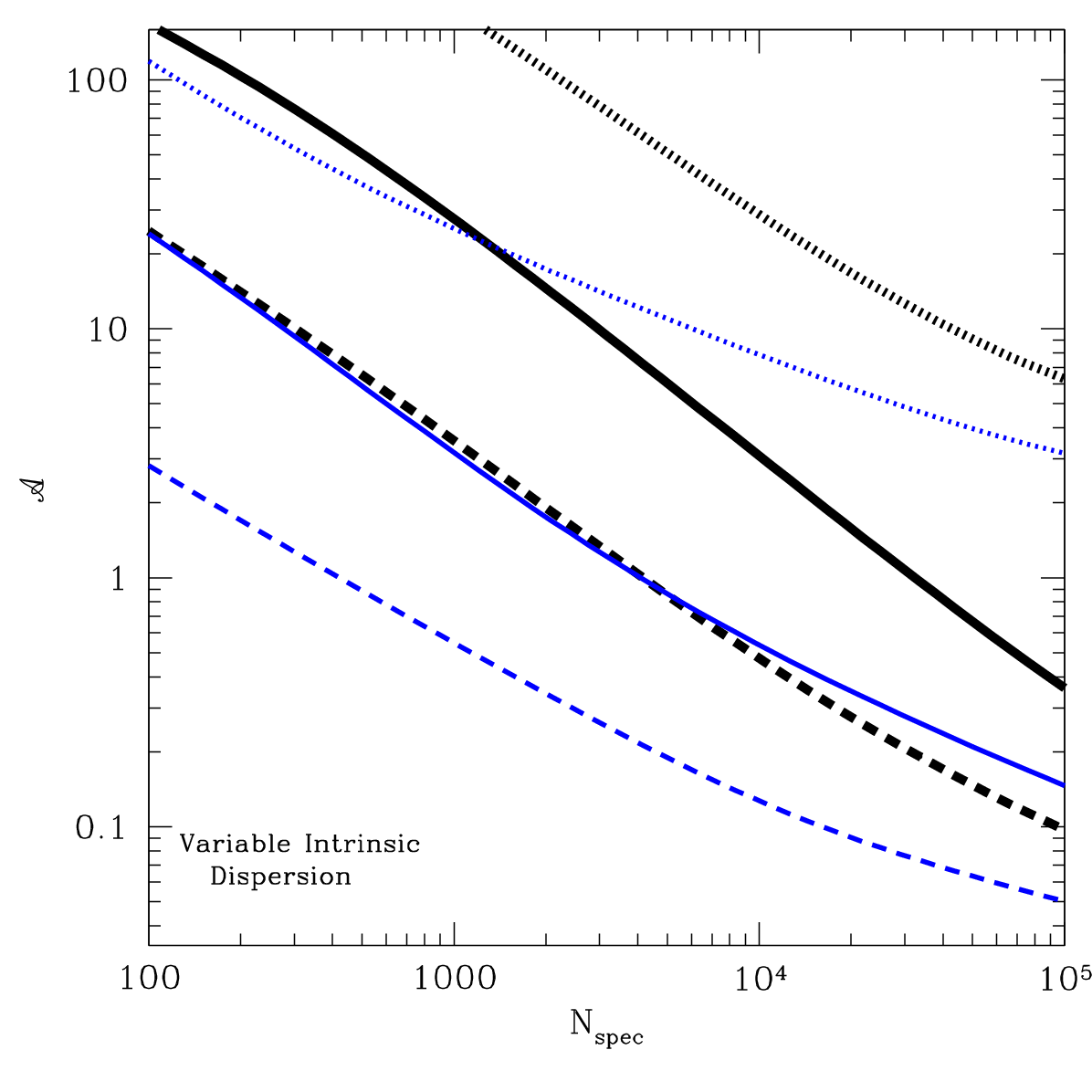}
\caption{
Influence of spectroscopic calibration sets on dark energy constraints 
from a large sample of SNeIa from a photometric survey.  This plot depicts the 
area of the $95\%$ contour in the $\wzero$-$\wa$ plane $\fom$, 
as a function of the size of the spectroscopic SNeIa sample used to 
calibrate the photometric redshift model, under several assumptions.  There are nine 
lines in the {\em left} panel.  The {\em solid} lines represent constraints as a function of 
spectroscopic sample size from the luminosity distance--redshift test only.  
The {\em dotted} lines represent constraints from the dispersion among SNeIa 
distance moduli.  The {\em dashed} lines show the total constraint from both 
the mean and dispersion in SNeIa distance moduli.  For each line type, the 
{\em thickest} lines show results in a standard case where the spectroscopic 
calibration sample traces the redshift distribution of SNeIa in 
the photometric survey.  Alternatively, the {\em thinnest} lines show $\fom$ 
under the assumption that the 
calibration sample is distributed uniformly in redshift over the interval 
$z=0.1-1.7$.  The lines of {\em intermediate} thickness show constraints 
with a spectroscopic sample of size $\nspec$ in addition to a low-redshift ``nearby'' 
spectroscopic sample of 500 SNeIa distributed uniformly in the interval 
$z=0.03-0.08$.  The {\em left} panel shows the case of a 
known and fixed intrinsic dispersion and the {\em right} panel shows the case 
of a variable intrinsic dispersion.  Note that in the case of a known intrinsic 
dispersion, the low-redshift, nearby SNe add little information to the dispersion 
only test, so the dispersion-only lines in these cases are nearly coincident.  
For clarity, in the {\em right} panel, 
we show only two cases, that of the standard sample and that of the standard 
sample with an additional, nearby spectroscopic sample. 
}
\label{fig:nspecfom}
\end{figure*}

There are several things to note in the left panel of Fig.~\ref{fig:w0wa}.  
First, in the limit of perfect knowledge of the photometric redshift 
distribution and perfect control of systematic uncertainties, 
constraints from the standard luminosity distance test 
dominate, placing a constraint on $\wpiv$ of $\sigma(\wpiv) \simeq 0.01$.  
However, the extra information contained in the SNeIa dispersions 
is not uninteresting.  The dispersion test constrains the pivot 
equation of state parameter to $\sigma(\wpiv) \simeq 0.07$ by itself.  
More importantly though, is that the luminosity distance and dispersion 
contain complementary information.  The luminosity distance test alone has a 
pivot redshift of $z \sim 0.1$, and so the constraint contours from this 
test alone are quite vertical.  The 
dispersion test has a pivot redshift at $z \sim 0.4$, and so the dispersion contours 
are more inclined and it is clear that the dispersion test complements the distance test.  
The combined constraint from the mean and dispersions among supernovae distance moduli 
result in a constraint on $\wpiv$ that is only slightly smaller than that from the 
distance test alone, but a significant reduction in the area of the 95\% contour 
by a factor of roughly $\sim 2$.  For reference, we also show constraints from 
weak lensing shear tomography for the LSST survey as estimated by \citet{zentner_etal08}, 
assuming photometric redshift calibration with a spectroscopic sample of $10^5$ galaxies 
as in {\tt Model I} of \citet{ma_etal06}.   
The weak lensing constraint on $\wpiv$ is $\sigma(\wpiv) = 0.013$, weaker than that from SNeIa 
in this idealistic scenario, but the area of the 95\% contour for weak lensing shear tomography 
is about 30\% smaller than that from SNeIa.  Aside from serving as a reference, this also 
demonstrates in a tangible way that the criteria of minimizing $\sigma(\wpiv)$ and $\fom$ 
can result in notably different conclusions.  In what remains, we make an effort 
to address constraints in a variety of more plausible scenarios.

The right panel of Fig.~\ref{fig:w0wa} depicts 1$\sigma$ constraint contours in 
the more realistic case of imperfect knowledge of the photometric redshift 
parameters.  As described in \S~\ref{section:methods}, we quantify the knowledge of 
the photometric redshift parameters in terms of the size of the spectroscopic sample 
used to place priors on these parameters, $\nspec$.  For the specific results in the right 
panel of Fig.~\ref{fig:w0wa}, we have chosen $\nspec=3000$ and assumed the spectroscopic calibration 
sample to be distributed uniformly in redshift over the range 
$0.1 \le z \le 1.7$, as might result from a forthcoming JDEM \citep[][]{kim_etal04,albrecht_etal06}.  
Note that the axes on the right panel of Fig.~\ref{fig:w0wa} span a different range from the left panel, 
but that the inner contour showing projected constraints from galaxy shear tomography are identical 
in each panel.

The right panel of Fig.~\ref{fig:w0wa} shows a considerable degradation of dark energy constraints 
from either the standard test of luminosity distance or the dispersion of SNeIa 
distance moduli alone relative to the case of perfect knowledge of the photometric 
redshift distribution.  This demonstrates the importance of both a realistic assessment of 
photometric redshift uncertainties and prior information 
to constrain photometric redshifts.  However, this panel also depicts the complementary 
nature of these constraints.  The luminosity distance test alone gives an error on the pivot equation 
of state parameter of $\sigma(\wpiv) \approx 0.09$.  
Utilizing the information contained within the dispersion 
of SNeIa as well as the mean decreases the error on the pivot equation 
of state parameter to roughly $\sigma(\wpiv) \simeq 0.06$ and decreases the 
area of the 95\% contours relative to either test individually by nearly a factor of ten by 
breaking the prominent degeneracy in the $\wzero$-$\wa$ parameter space.  
Worthy of note is the fact that at this level, the constraints from the large, 
photometric SNeIa sample are comparable to the constraints 
expected from the spectroscopic sample used for photometric redshift calibration 
\citep[see also][]{albrecht_etal06}, so they are both interesting and relevant, but the 
availability of a significantly larger spectroscopic sample would obviously be more 
valuable because of the constraints it can place on dark energy directly, rather 
than its ability to calibrate the redshifts of a photometric sample.

Though Fig.~\ref{fig:w0wa} shows contours in two simple models, the basic point is 
that SNeIa dispersions derived from forthcoming large, photometric surveys are potentially useful.  
Generally, the luminosity distance test is subject to a strong degeneracy between 
$\wzero$, $\wa$, and $\Omegam$.  Our choice of priors mitigates the influence of 
$\Omegam$, but the $\wzero$-$\wa$ degeneracy remains because the sample has only a 
relatively small redshift span and therefore limited leverage with which to 
measure any time variation of the dark energy equation of state.  Some contemporary studies 
and several future proposals for SNeIa-based dark energy experiments (such as a 
Joint Dark Energy Mission like SNAP) aim to break this degeneracy by observing 
high-redshift survey \citep[e.g.,][]{riess_etal07}.  On the contrary, 
our assumed photometric survey has no high-redshift component.  In fact, the number of 
SNeIa in our assumed survey declines rapidly for $z \gsim 0.6$ and only a few percent of 
the SNeIa in the survey have $z \gsim 0.8$ \citep[see][]{zhan_etal08}.  Within the photometric survey, the 
$\wzero$-$\wa$ degeneracy is broken by the complementarity between the 
dispersion due to gravitational lensing and the standard luminosity distance test.
This is a potential aspect of complementarity between spectroscopic SNeIa samples 
and samples from a large, ground-based, photometric survey.  
The latter, bereft of high-redshift ($z \gsim 0.8$) 
objects, will be sensitive to $w(a)$ at lower redshifts than proposed spectroscopic 
SNeIa projects that typically extend to $z \gsim 1.5$.  Moreover, 
the dispersion test probes the inhomogeneities in the universe.  As such, 
the dispersion test probes a fundamentally different manifestation of dark energy, 
namely its influence on cosmic structure growth, that is not probed by the 
luminosity distance test.

A further benefit of the additional dispersion information is that it helps to 
break degeneracies between cosmological and photometric redshift parameters.  In fact, 
when only considering the canonical luminosity distance test, the only constraints on the 
photometric redshift parameters come from the priors determined by the photometric 
redshift calibration set.  When both pieces of information are exploited, the data 
provide for weak ``self-calibration,'' determining many of the photometric redshift 
parameters with uncertainties a factor of $\sim 2-5$ more stringently than the priors 
alone.  Moreover, Fig.~\ref{fig:w0wa} also shows that the dispersion 
test is less sensitive to photometric redshift uncertainties and degrades more 
slowly with decreasing knowledge of the photometric redshift distribution.  This useful 
and interesting result stems primarily from the fact that the lensing kernel 
[$W(z',z)$ in Eq.~(\ref{eq:lens})] is inherently 
broad and the dispersion is a comparably mild function of redshift.

As {\em some} spectroscopic calibration of photometric redshifts is necessary 
in order to utilize a photometric SNeIa sample, it is 
interesting to consider potential constraints both as a function of 
the uncertainty in photometric redshift parameters and under a variety of assumptions 
about what additional SNeIa data will be available.  This permits an estimate of the 
utility of SNeIa from a future large-scale, photometric survey given specific photometric 
redshift calibration programs and additional calibration data sets.  For this reason, 
the final two plots of this section present results for dark energy parameter constraints 
as a function of spectroscopic SNeIa calibration sample size $\nspec$, under a variety 
of assumptions about additional available data.

\begin{figure*}[t!]
\includegraphics[width=8.0cm]{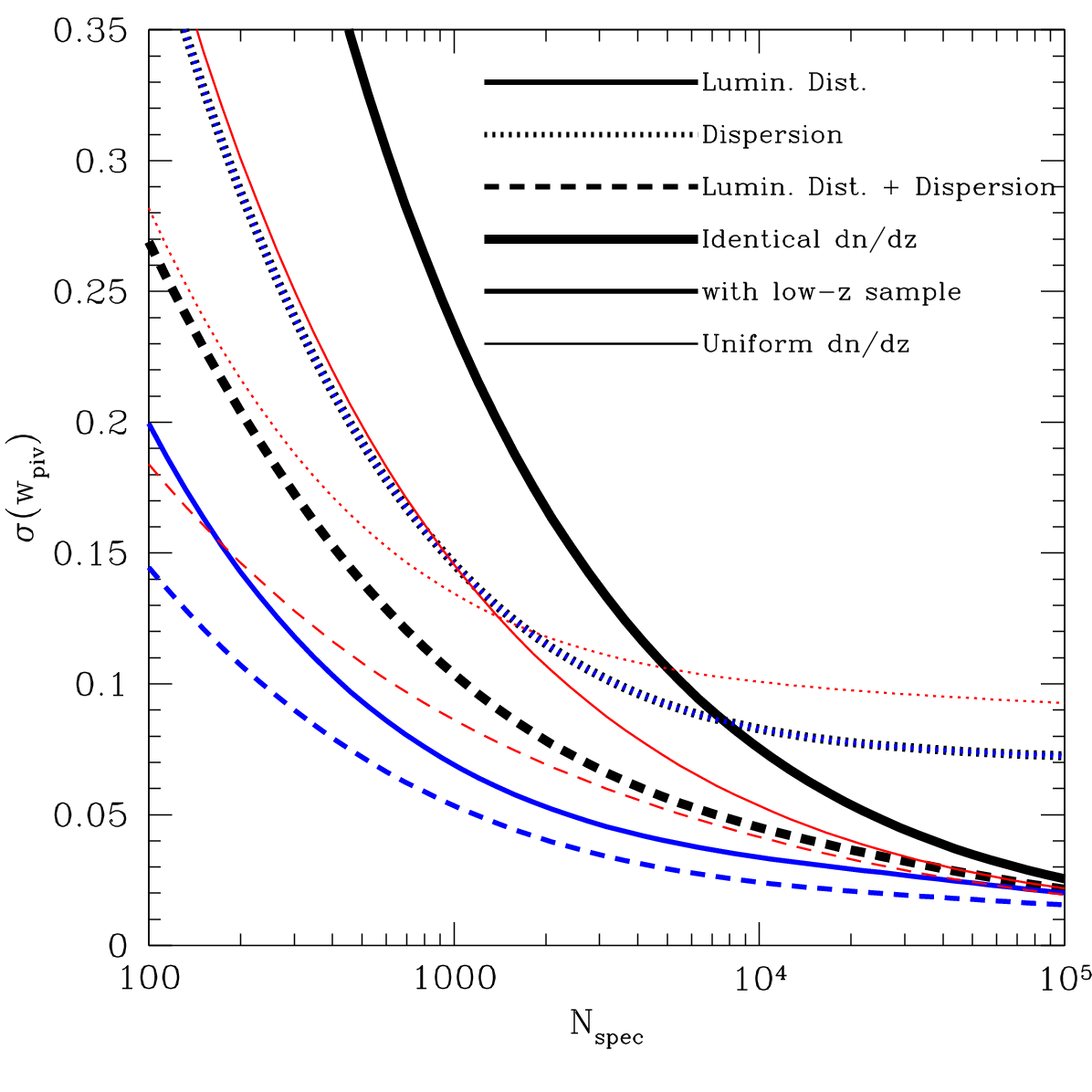}
\includegraphics[width=8.0cm]{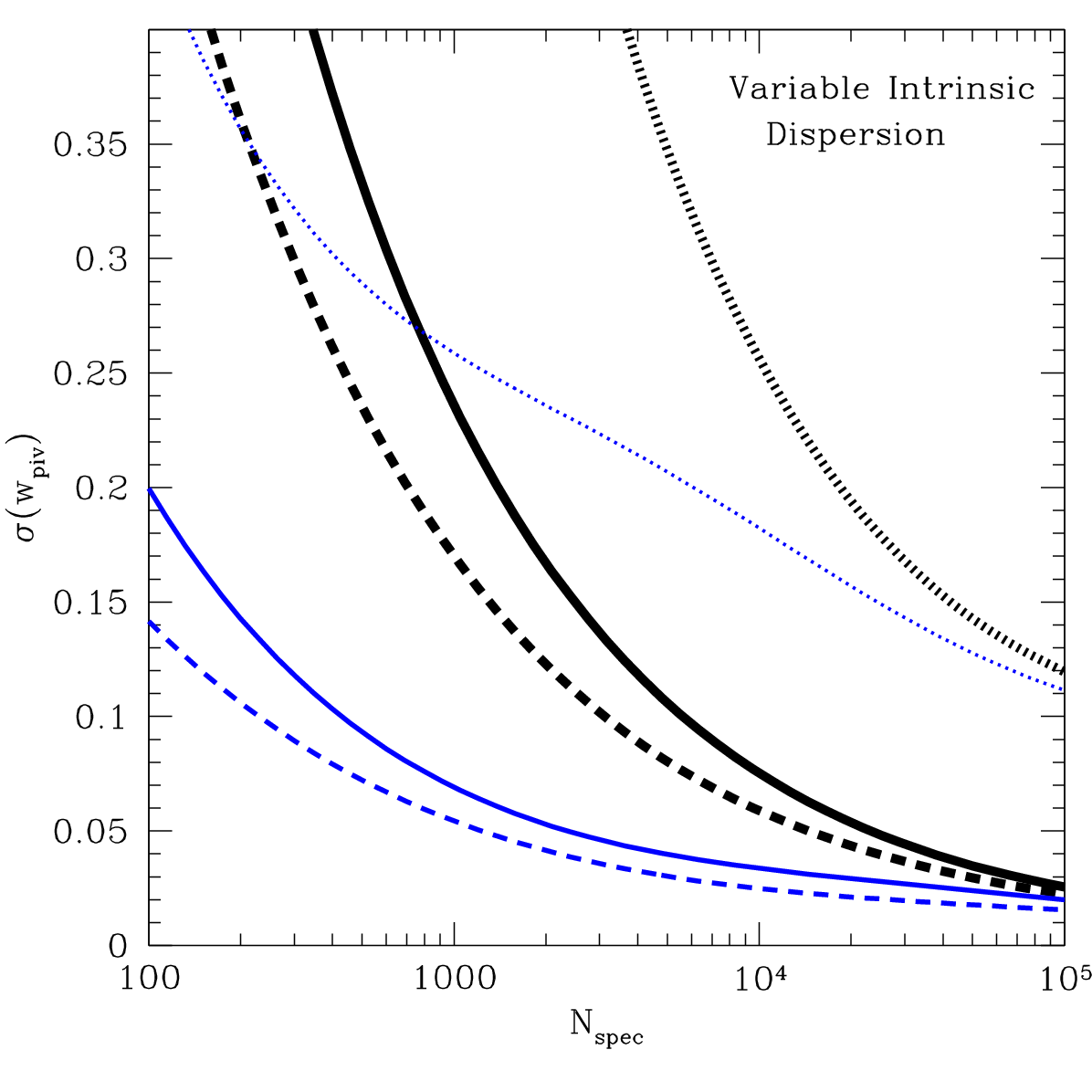}
\caption{
Constraints on the dark energy pivot equation of state from a large sample of 
SNeIa from a photometric survey as a function of the size of the 
spectroscopically-observed SNeIa set available to calibrate photometric redshifts.  
This plot depicts constraint on $\wpiv$ as a function of the 
size of the spectroscopic sample used $\nspec$, under several assumptions.  There are nine 
lines in the {\em left} panel.  The {\em solid} lines represent constraints as a function of 
spectroscopic sample size from the luminosity distance--redshift test only.  
The {\em dotted} lines represent constraints from the dispersion among SNeIa 
distance moduli.  The {\em dashed} lines show the total constraint from both 
the mean and dispersion in SNeIa distance moduli.  For each line type, the 
{\em thickest} lines show the results from a standard case where the 
redshift distribution of the spectroscopic 
calibration sample traces the redshift distribution of SNeIa in 
the imaging survey.  The {\em thinnest} lines show $\sigma(\wpiv)$ assuming 
that the calibration sample is distributed uniformly in redshift across the interval 
$z=0.1-1.7$.  The lines of {\em intermediate} thickness show constraints 
with a spectroscopic sample of size $\nspec$ in addition to a low-redshift ``nearby'' 
spectroscopic sample of 500 SNeIa distributed uniformly over the interval 
$z=0.03-0.08$.  The {\em left} panel shows the case of a 
known and fixed intrinsic dispersion and the {\em right} panel shows the case 
of a variable intrinsic dispersion.  
Note that in the case of a known intrinsic 
dispersion, the low-redshift, nearby SNe add little information to the dispersion 
only test, so the dispersion-only lines in these cases are nearly coincident.  
For clarity, in the {\em right} panel, 
we show only two cases, that of the standard sample and that of the standard 
sample with an additional, nearby spectroscopic sample.
}
\label{fig:nspecwpiv}
\end{figure*}

First, we consider the area of the 95\% ellipse in the $\wzero$-$\wa$ plane $\fom$, 
as a function of spectroscopic calibration sample size, $\nspec$.  This quantity is 
shown in Figure~\ref{fig:nspecfom} for several different combinations of data and 
model assumptions.  In the left panel of Fig.~\ref{fig:nspecfom}, we display results for 
models in which the intrinsic dispersion is set to a fixed value and in the right panel, 
for models in which the intrinsic dispersion level is free to float.  
The thickest lines in Fig.~\ref{fig:nspecfom} show our standard model constraints 
with no nearby SNeIa sample and a calibration set distributed in redshift 
in a manner identical to the SNeIa in the photometric sample.  The intermediate thickness 
lines show $\fom$ as a function of $\nspec$ in a model that includes the nearby sample 
of SNeIa.  The thinnest lines in Fig.~\ref{fig:nspecfom} represent the standard model 
with a calibration set that is distributed uniformly in redshift in the interval 
$0.1 \le z \le 1.7$.  For illustrative purposes, we show constraints for 
spectroscopic sample sizes up to $\nspec = 10^5$; however, we note that such 
high $\nspec$ are of limited practical interest because an upper limit on 
achievable spectroscopic sample sizes is $\nspec \sim 10^4$ and at such large $\nspec$ 
constraints from the spectroscopic sample itself begin to dominate any 
added information from the photometric sample.  Note that in Fig.~\ref{fig:nspecfom} 
the lines that delineate constraints from the dispersion test only in the standard 
case and in the case with an additional nearby SNeIa sample are nearly coincident, 
and may not be readily discernible.  The reason for this is simple.  Under the 
assumption that the dispersion is known, the nearby sample does not serve to 
calibrate the intrinsic dispersion and adds little information.

The complementarity of the information contained in the mean, $\meanmui$, and 
the dispersion, $\sigmui$, is apparent in Fig.~\ref{fig:nspecfom}.  The information 
contained in SNeIa dispersions is not negligible and in some cases the constraints 
from SNeIa dispersions can exceed those from the luminosity distance test alone in 
the regime $\nspec \lsim 10^4$.  The most interesting cases 
are likely those where the intrinsic dispersion is not known at the outset and/or 
where there is a nearby SNeIa sample available.  In these cases, 
the information from the standard luminosity distance test alone yields a significantly 
lower area $\fom$, than the information from the dispersion among distance moduli.  
Combining constraints from both the mean and the dispersion information 
drives $\fom$ down by roughly a factor of ten compared to either alone in all cases.  This is 
a manifestation of the fact that these two observables are most effective probes at different 
redshifts and so exhibit different degeneracies in the $\wzero$-$\wa$ plane 
(see Fig.~\ref{fig:w0wa}).  In addition, Fig.~\ref{fig:nspecfom} illustrates 
that constraints from $\sigmui$ degrade more slowly with decreasing $\nspec$ than 
do constraints from the standard luminosity distance test, a feature already 
evident in Fig.~\ref{fig:w0wa}.

In general, including the low-redshift, nearby SNe sample 
drives constraints considerably lower \citep[][]{aldering_etal02,copin_etal06,albrecht_etal06}, 
greatly increasing the utility of the luminosity 
distance test while only yielding a slight improvement on the dispersion test {\em if} 
the intrinsic dispersion level is assumed to be known.  Assuming a calibration set of 
spectroscopically-observed SNeIa that is distributed uniformly in redshift is moderately 
more powerful than the calibration set that follows the redshift distribution of the SNeIa 
in the photometric sample.  This is because the uniform redshift distribution allows for 
more accurate photometric redshift calibration at high $\zp$.  This is important for two reasons.  
First, deviations between observables predicted by different models are larger at high redshift.  
Second, the redshift distribution of SNeIa $\dd n/\dd z$ is a steep function of redshift, 
dropping as $\dd n/\dd z \propto \exp[-32(z-0.5)^2]$ at $z \gg 0.5$, so the scatter of 
SNeIa in this regime to higher redshift due to redshift errors can lead to significant 
contamination and a Malmquist-type bias.  In all cases, $\fom$ continues to decline rapidly with $\nspec$.  In 
principle, this rate of decline levels off as knowledge of photometric redshift 
distributions become so accurate as to allow for the recovery of all 
available cosmological information.  This drives all of the constraints to begin to converge 
at large $\nspec$.  In practice, this saturation is not achieved until $\nspec \sim 3 \times 10^6$, 
a limit that is so large as to be uninteresting (it is larger than the photometric SNeIa sample!).

The right panel differs from the left in that the intrinsic dispersion level $\sigint$, 
is treated as unknown and permitted to vary along with $\Delta$, the cosmological parameters, and the 
photometric redshift parameters.  This leads to considerable degradation in the 
utility of the dispersion alone to constrain dark energy, but in this case 
the dispersion information constrains the intrinsic 
dispersion, determining it to $\sigma(\sigint) \simeq 1.6 \times 10^{-3}$.  For comparison, 
we have evaluated the ability of either the nearby sample or a spectroscopic sample with 
with $3000$ SNeIa spread uniformly over the redshift interval $0.1 \le z \le 1.7$ 
and redshift errors of $\sigma_z=10^{-4}$ to constrain $\sigint$.  
We find that determinations from such data are similar to, but slightly less restrictive, 
than that from the photometric survey after marginalizing over all other parameters.  
Conversely, much of the loss of information due to a varying level of intrinsic dispersion 
is largely {\em recovered} in the case where the nearby SNeIa sample is included.  
Notice that complementarity between the information in $\meanmui$ and $\sigmui$ 
leads to considerable improvements in $\fom$ in all cases.

Another useful measure of the constraining power of any set of observables is the error on the 
equation of state parameter at the pivot redshift, $\sigma(\wpiv)$.  We show the dependence 
of $\sigma(\wpiv)$ upon $\nspec$ in Figure~\ref{fig:nspecwpiv}.  First, note that $\sigma(\wpiv)$ 
varies over a smaller dynamic range than $\fom$, so Fig.~\ref{fig:nspecwpiv} 
has an ordinal axis with a linear scale as opposed to the logarithmic scale used in Fig.~\ref{fig:nspecfom}.  
After accounting for this difference there are yet some 
qualitative differences in the behavior of $\sigma(\wpiv)$ relative to $\fom$.  
In the limit of small $\nspec$ and in the absence of a large, low-redshift, nearby 
sample of SNeIa the dispersion information can give useful, independent constraints on $\wpiv$.  
Moreover, the dispersion information is generally less useful for improving $\sigma(\wpiv)$ 
than it is for improving the $\fom$ figure of merit, particularly when the photometric redshift 
parameters are very well known or the nearby SNeIa sample is available.  In these 
cases, $\fom$ improves because the error ellipses or oriented at a significant angle 
relative to each other, but the shortest axis (the constraint on $\wpiv$) 
is largely determined by the luminosity distance test alone.  Lastly, it is worthwhile noting that 
in the case where there is an available nearby spectroscopic sample, 
the marginal improvement in $\wpiv$ upon adding additional objects 
to the spectroscopic photometric calibration sample is 
relatively small beyond $\nspec \sim 2 \times 10^3$.

To this point, we have considered cases in which the intrinsic dispersion of SNeIa does not 
vary with redshift.  It is not unlikely that the intrinsic dispersions in SNeIa will change with redshift.  
For example, SNeIa will be observed in different bands relative to their rest frames 
at different redshifts with the consequence that standard candle calibration will be a function 
of redshift \citep[e.g.,][]{hamuy_etal95,prieto_etal06}.  In addition, there is evidence that the 
Ia class of supernovae is composed of distinct sub-classes and the relative mix of these sub-classes 
is expected to change with redshift \citep{hamuy_etal95,howell01,howell_etal07}.  
Clearly, if the intrinsic dispersion is allowed to be an {\em arbitrary} function of redshift, the utility 
of the added dispersion information is completely eliminated as a cosmological probe.  
However, we have considered a less pathological case of a 
redshift-dependent intrinsic dispersion that is monotonic and follows a power-law.

To be specific, we considered a model with $\sigint(z)=\sigizero(1+z)^{\beta}$ with both 
$\sigizero$ and $\beta$ parameters that are fixed by the data ($\beta=0$ in the fiducial model 
about which we perturb).  We do not show the 
results in Fig.~\ref{fig:nspecfom} or Fig.~\ref{fig:nspecwpiv} for 
clarity, but note that further degradation is not devastating.  
In particular, in the case with no nearby sample, further degradation 
by allowing for $\sigint(z)$ to vary as a power law in $(1+z)$ is a factor of 
$\sim 2.5$ in $\fom$ and $\sim 18\%$ in $\sigma(\wpiv)$ relative to the case of 
constant $\sigint$.  With a nearby spectroscopic sample, 
$\sigint(z)$ at low redshift can be effectively determined and the degradation relative to 
the case of a redshift-independent intrinsic dispersion is only $\sim 60\%$ in $\fom$ and 
just $\sim 10\%$ in $\sigma(\wpiv)$.  In addition, these data would limit $\beta$ with 
uncertainties of $\sigma(\beta) \simeq 0.22$ and $\sigma(\beta) \simeq 0.09$ in cases with 
and without the nearby sample respectively, a level that is again comparable to the 
forecasts from a spectroscopic sample.  Moreover, it is clear that the lensing contribution 
to the dispersion must be accounted for in order to extract the intrinsic SNeIa properties.  
In summary, though the information loss in cases 
of redshift-dependent intrinsic dispersion is not insignificant, SNeIa in a large, photometric survey 
can bring significant constraining power to bear on dark energy even with moderate, and 
poorly-understood time variation of the intrinsic dispersion.  Of course, to the degree that 
the distribution of distance moduli varies markedly and rapidly 
with redshift outside of the expected range, this distribution 
will provide a useful handle on the SNeIa properties themselves, including 
perhaps the evolution of SNeIa and SNeIa population demographics 
\citep[e.g.,][]{hamuy_etal95,howell01,mannucci_etal06,sullivan_etal06,howell_etal07,sarkar_etal08b}.

\section{Implications and Caveats}
\label{section:discussion}

We have studied simultaneous constraints on dark energy 
coming from both the evolution of the mean 
luminosity distance of SNeIa as a function of redshift and 
the dispersion among SNeIa fluxes in a SNeIa 
data set that may arise from a large, wide, and fast photometric survey such 
as that proposed for the LSST.  Sources of dispersion among the measured fluxes of 
SNeIa in such a survey include intrinsic disperision among SNeIa (including 
standard candle calibration), uncertainty in photometric redshifts, and 
magnification due to gravitational lensing.  We have shown that the additional 
dispersion information complements the traditional luminosity 
distance test in several ways.

First, the dispersion information breaks 
a degeneracy between the contemporary dark energy equation of state $\wzero$, 
and dark energy equation of state evolution, 
parameterized here by the common, benchmark parameter $\wa$.  Using the luminosity 
distance (mean flux) test alone, this degeneracy can be broken by observing 
high-redshift SNeIa to increase the lever arm of the data in redshift.  
The dispersion information leads to constraints that break the degeneracy between 
$\wzero$ and $\wa$ without the nead for a high-redshift ($z \gsim 0.8$) 
SNeIa sample (see Fig.~\ref{fig:w0wa}).

Second, photometric surveys require some calibration of photometric redshifts.  
Including dispersion information may allow for mild internal 
self calibration of uncertainties in the true redshift distribution 
of observed SNeIa given their photometric redshifts.  In the context of our models, 
this results in constraints on photometric redshift model parameters 
(see \S~\ref{section:methods}) that are a factor of $\sim 2-5$ more constraining 
than the priors on these parameters from a spectroscopic sample of size $\nspec=3000$.  
The realized level of improved calibration depends upon the details 
of both the model and the additional SNeIa calibration data that are available, 
but in all cases the luminosity distance test alone does not serve to calibrate 
the photometric redshift distribution.

Lastly, it is entirely possible that there will be evolution in the distribution 
of SNeIa intrinsic luminosities that significantly degrade the forecasts for dark 
energy constraints that we present (see \S~\ref{section:results}).  
All approaches to SNeIa cosmology, even those that 
exploit the traditional luminosity distance test alone, rely upon some knowledge and/or 
assumptions about the intrinsic distribution of SNeIa luminosities to obtain cosmological 
constraints.  In cases where the evolution of the spread in luminosities at fixed redshift 
is significant over the redshift range $0.2 \lsim z \lsim 0.8$, a photometric survey 
cannot use lensing to constrain dark energy directly.  However, 
we have shown that such a survey can measure the evolution of the 
intrinsic SNeIa luminosity distribution at levels comparable to, or better than, 
a spectroscopic survey containing a few thousand SNeIa.  This should aid the 
luminosity distance test by providing a better understanding of possible SNeIa 
evolution and demographics and, therefore, 
a better understanding of potential parameter biases and uncertainties.

As we have mentioned in the preceding paragraph, 
in examining a photometric SNeIa sample, it is necessary to consider 
uncertainties in the photometric redshift distribution.  We have assumed that 
some calibration will be done with a spectroscopic sample of an unspecified size, $\nspec$.  
Figure~\ref{fig:nspecfom} gives the expected area of the 95\% ellipse in the $\wzero$-$\wa$ 
plane, $\fom$, as a function of $\nspec$.  $\fom$ decreases rapidly with 
$\nspec$ in all cases because increasing $\nspec$ allows access to information over a 
broader range of observed redshifts and helps to break degeneracies in the $\wzero$-$\wa$ 
plane.  Figure~\ref{fig:nspecwpiv} shows the constraints on the equation of state 
parameter at that redshift where the data best constrain it, the so-called pivot 
equation of state $\wpiv$, as a function of $\nspec$.  This figure shows that the 
marginal improvement in $\sigma(\wpiv)$ with increasing $\nspec$ decreases
beyond $\nspec \sim \mathrm{a}\ \mathrm{few}\ \times 10^3$.  This result is interesting 
because the upper limit on the achievable size of a spectroscopic sample is near $\nspec \sim 10^4$ 
and, moreover, with $\nspec \gsim 10^4$ constraints from the spectroscopic sample alone 
(which has miniscule redshift errors, $\sigma_z \sim 10^{-4}$) would begin to 
dominate over that achievable with a large, photometric sample even in the ideal limit 
where statistics, and not systematics, dominate the error budget.  Both Fig.~\ref{fig:nspecfom} 
and Fig.~\ref{fig:nspecwpiv} illustrate that at fixed $\nspec$ 
it is more fruiful to employ a spectroscopic calibration set 
weighted more toward high-redshift ($z \gsim 0.5$) SNeIa than the 
distribution of SNeIa from the imaging survey.

The previous paragraph closed with an allusion to systematic error, and indeed 
sources of systematic error are a major caveat to the present study.  The 
main advantage of a photometric sample is its size and the additional statistical 
leverage that come with a large sample size.  However, once systematic errors dominate 
the error budget, it is no longer possible to take advantage of this benefit.  
What is more, a photometric survey places emphasis on many potential sources of systematic 
error.  We have accounted for some of these complications.  For example, we have 
allowed for consistent determination of the photometric redshift distribution 
and for variation of the average SNeIa luminosity and intrinsic dispersion, 
including mild evolution.  This is indicative that our results will be 
somewhat robust to evolution in the intrinsic SNeIa population 
\citep[e.g.][]{hamuy_etal95,howell01,sullivan_etal06,howell_etal07,sarkar_etal08b}.  
Other sources of systematic error are likely to be important.  In fact, the 
SNAP collaboration, which aims to observe and exploit a 
spectroscopically-observed SNeIa sample from space, 
has assumed a fundamental limit in the ability to calibrate 
supernovae to within $\delta \mu \sim 0.02$~mag over a redshift difference of 
$\delta z \sim 0.1$ to specify their survey \citep[e.g.][]{kim_etal04}.  
Yet it remains to be seen whether such a systematic limit will be realized, and 
it is possible that it may be higher or lower.  
Further, without spectroscopy, not only are redshifts uncertain, but 
supernovae type identification, extinction and K-corrections, and 
standard candle calibration all become more difficult 
\citep[e.g.][]{pinto_etal04,prieto_etal06}.  We have not 
explicitly accounted for any of these, largely because systematic 
levels are uncertain and modeling them is both complex and dependent upon 
survey strategy.  These complications place a detailed treatment of systematics 
beyond the scope of our study (the studies by the DETF, as well as those of 
\citealt{zhan_etal08} and \citealt{hannestad_etal08} neglected a detailed 
treatment of systematics as well).  
If the assumed SNAP systematics floor is realized, 
our forecasts for dark energy equation of state constraints 
increase by a factor of $\sim 3$, significantly reducing the additional cosmological information 
that may be extracted from the photometric sample of SNeIa.

\begin{figure*}[t!]
\includegraphics[width=8.0cm]{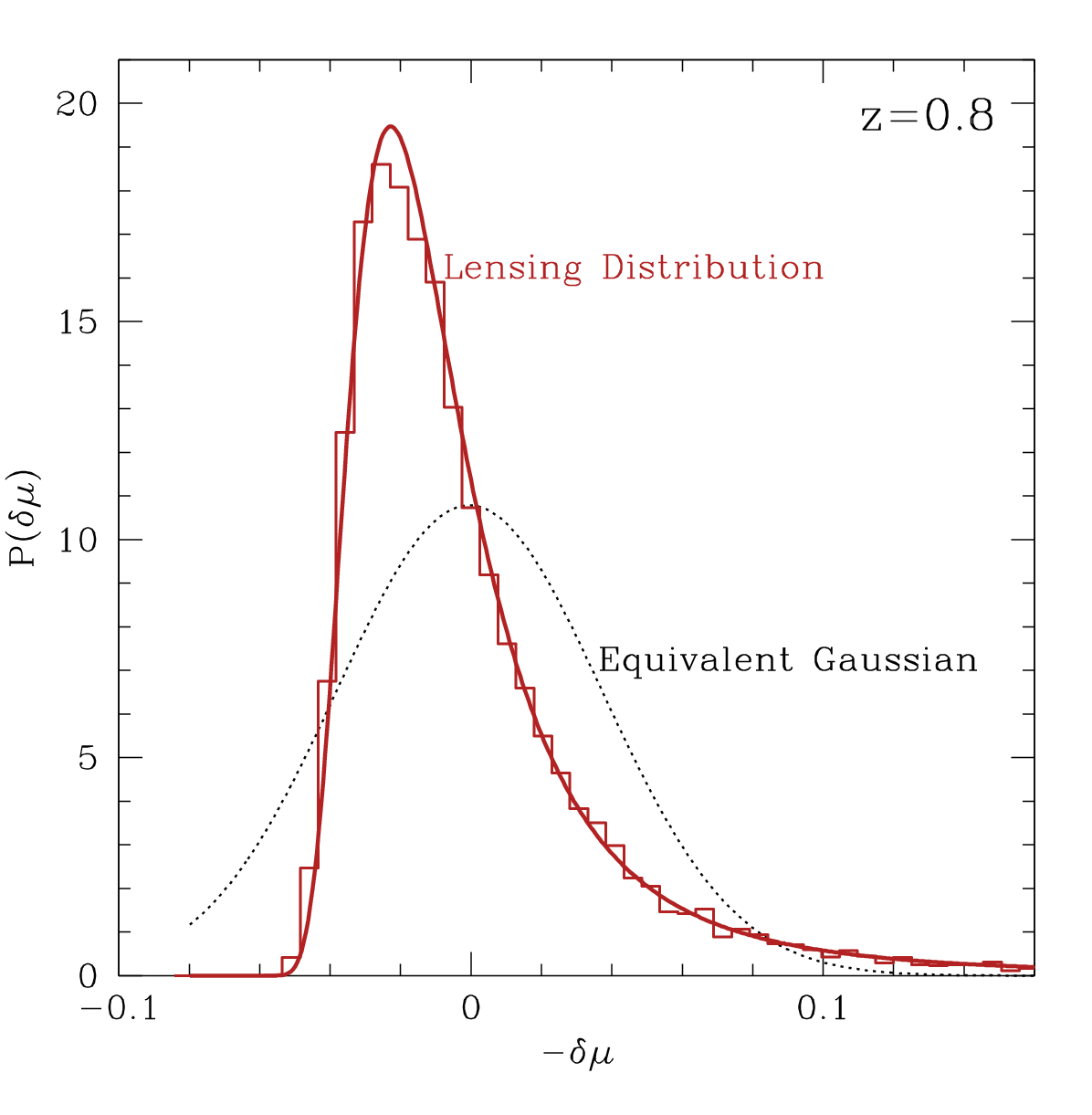}
\includegraphics[width=8.0cm]{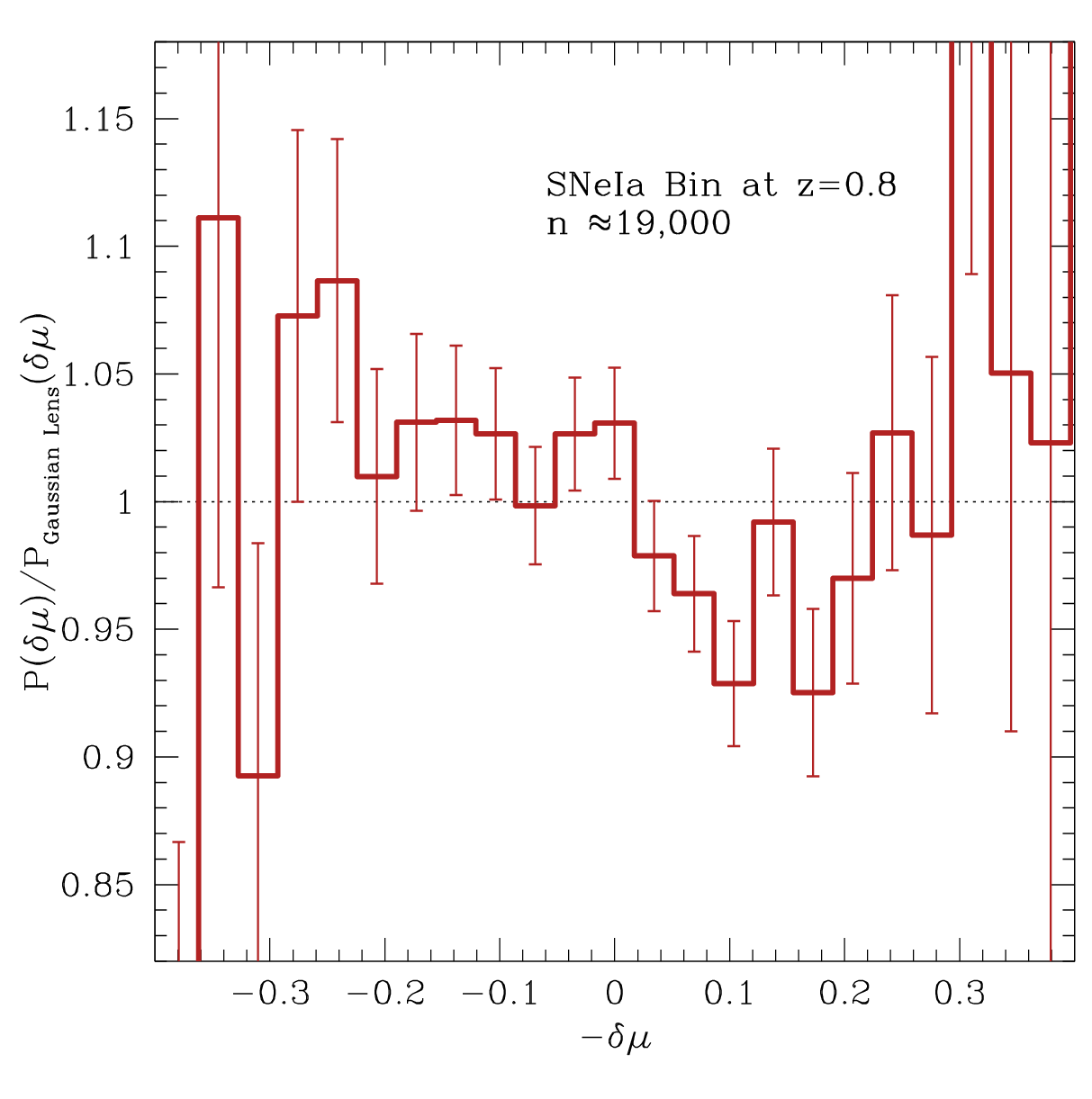}
\caption{
The non-Gaussianity of the lensing distribution.   Both panels show the 
distribution of the shift in distance modulus $\delta \mu$ of SNeIa.  The 
{\em left} panel is illustrative.  The {\em solid, smooth} curve shows the 
distribution of distance moduli for sources at redshift $z=0.8$ due only to 
lensing.  This result was computed using the fitting 
formula of \citet{wang_etal02}.  We display this 
quantity at relatively high redshifts only because the formula of 
\citet{wang_etal02} is no longer internally self-consistent below $z \sim 0.6$.  
The {\em dotted} curve shows a Gaussian distribution with the same mean 
and dispersion as the lensing distribution.  The Gaussian distribution is 
truncated at the minimum magnification, corresponding to a maximum shift in 
distance modulus $\delta \mu_{\mathrm{max}}$.  The histogram shows the distribution 
of distance modulus shifts from a random realization of $19,000$ SNeIa at $z=0.8$.  
The {\em right} panel shows something closer to what would be seen in any 
observed set of SNeIa.  In this panel, we show the distribution of distance moduli 
(relative to the mean) for a random sample of SNeIa in a bin of width 
$\delta z = 0.03$ at $z=0.8$.  The sample consists of $19,000$ SNeIa, which 
is roughly the number of SNeIa expected in such a bin according to our 
fiducial model described in \S~\ref{section:methods}.  In this 
case, the distribution includes the intrinsic dispersion, the dispersion caused by 
poorly-controlled photometric redshifts, and the lensing distribution.  For 
clarity, the probability distribution is plotted relative to the same distribution 
under the assumption of the lensing contribution is a Gaussian, 
$P_{\mathrm{Gaussian\ Lens}}(\delta \mu)$.  
The error bars reflect the statistical error on $P(\delta \mu)$ from a sample 
of this size and portray of the detectability of the 
unique shape induced by lensing in the absence of significant errors.
}
\label{fig:plens}
\end{figure*}

As described in \S~\ref{section:methods}, the calculations we have performed 
to estimate the additional information contained in the variety of 
SNeIa distance moduli at a fixed redshift are greatly simplified.  
Our idealization may lead to an over-estimate of 
the constraining power of SNeIa.  However, we have also 
neglected some potentially-available information.  In particular, the distribution 
of lensing magnifications is highly non-Gaussian 
\citep[][]{schneider_wagoner87,sasaki87,wambsganss_etal97,munshi_jain00,wang_etal02} 
and we have neglected any information beyond the dispersion because including the 
full information is computationally-intensive and current theoretical estimates 
of the magnification distribution are inadequate for our application. The left panel of 
Figure~\ref{fig:plens} shows the non-Gaussianity 
of the lensing distribution at a fixed redshift $z=0.8$ using the 
fitting formula of \citet{wang_etal02}.  The right panel of 
Figure~\ref{fig:plens} shows the distribution of distance moduli in a 
bin of width $\delta z = 0.03$ centered at redshift $z=0.8$.  In our fiducial 
model, the bin contains roughly $n \approx 19,000$ SNeIa and Fig.~\ref{fig:plens} 
demonstrates that the non-Gaussianity of the lensing distribution would be 
detectable even by examining the distribution of distance moduli in this single bin.  
For the realization shown, the non-Gaussianity of the lensing distribution is 
detectable at slightly more than $\sim 2\sigma$ from this bin alone.  
This strongly suggests that the lensing deviation from Gaussianity may 
be detectable in a complete analysis of a future photometric survey, where the full 
shape of the magnification distribution would be utilized and where all supernovae 
could be considered (which requires better predictions).  This is important for two reasons.  
First, the unique distribution of fluxes induced by 
lensing may aid in distinguishing the gravitational lensing effect from other additional 
sources of dispersion and from evolution in the intrinsic properties of SNeIa.  Second, 
to the degree that this non-Gaussianity can be detected, it will bring additional 
information to bear and improve dark energy constraints.

To conclude, a future imaging survey that scans a large fraction of the sky 
rapidly will enable the discovery of $\sim 10^5 - 10^6$ SNeIa.  The survey of the 
LSST is the canonical example of such an endeavor.  The number of 
SNeIa with light curves of sufficient quality to be used for cosmology depend 
upon specific survey strategies; however, such a SNeIa survey may be able 
to constrain dark energy properties using both the traditional luminosity 
distance test and the spread of supernovae apparent brightnesses as a function 
of photometric redshift.  Unfortunately, current theoretical treatments of 
gravitational lensing, in particular estimates of the magnification distribution, 
are not reliable enough to analyze any such data set.  In the coming years, it 
will be necessary to refine these estimates, including potential theoretical 
uncertainties \citep{huterer_takada05,zhan_knox04,white04,rudd_etal08}.  
In the end, the utility of such a sample to constrain 
dark energy is subject to limitations of systematic error and SNeIa 
evolution.  However, if observational systematics or SNeIa evolution 
prove to be limiting, such a large photometric survey will provide useful 
information about these issues.  In either case, an LSST-like imaging survey 
will revolutionize SNeIa cosmology.

\acknowledgments

We would like to thank Lloyd Braun, David Cinabro, Sourish Dutta, Scott Dodelson, 
Dragan Huterer, Arthur Kosowsky, Jeff Newman, 
Alberto Vallinotto, and Michael Wood-Vasey for useful discussions 
and email exchanges during the course of this work.  
ARZ is supported by the University of Pittsburgh and by the National Science Foundation through grant 
NSF AST 0806367.  ARZ thanks the Michigan Center for Theoretical Physics at the University of Michigan 
for hospitality and support while some of this work was performed.  
SB is funded by a Mellon Predoctoral Fellowship at the University of Pittsburgh.  
This research made use of the National Aeronautics and Space 
Administration Astrophysics Data System.

\bibliography{ms}

\begin{thebibliography}{59}
\expandafter\ifx\csname natexlab\endcsname\relax\def\natexlab#1{#1}\fi

\bibitem[{{Albrecht} {et~al.}(2006){Albrecht}, {Bernstein}, {Cahn}, {Freedman},
  {Hewitt}, {Hu}, {Huth}, {Kamionkowski}, {Kolb}, {Knox}, {Mather}, {Staggs},
  \& {Suntzeff}}]{albrecht_etal06}
{Albrecht}, A., {Bernstein}, G., {Cahn}, R., {Freedman}, W.~L., {Hewitt}, J.,
  {Hu}, W., {Huth}, J., {Kamionkowski}, M., {Kolb}, E.~W., {Knox}, L.,
  {Mather}, J.~C., {Staggs}, S., \& {Suntzeff}, N.~B. 2006, (astro-ph/0609591)

\bibitem[{{Aldering} {et~al.}(2002){Aldering}, {Adam}, {Antilogus}, {Astier},
  {Bacon}, {Bongard}, {Bonnaud}, {Copin}, {Hardin}, {Henault}, {Howell},
  {Lemonnier}, {Levy}, {Loken}, {Nugent}, {Pain}, {Pecontal}, {Pecontal},
  {Perlmutter}, {Quimby}, {Schahmaneche}, {Smadja}, \&
  {Wood-Vasey}}]{aldering_etal02}
{Aldering}, G., {Adam}, G., {Antilogus}, P., {Astier}, P., {Bacon}, R.,
  {Bongard}, S., {Bonnaud}, C., {Copin}, Y., {Hardin}, D., {Henault}, F.,
  {Howell}, D.~A., {Lemonnier}, J.-P., {Levy}, J.-M., {Loken}, S.~C., {Nugent},
  P.~E., {Pain}, R., {Pecontal}, A., {Pecontal}, E., {Perlmutter}, S.,
  {Quimby}, R.~M., {Schahmaneche}, K., {Smadja}, G., \& {Wood-Vasey}, W.~M.
  2002, in Presented at the Society of Photo-Optical Instrumentation Engineers
  (SPIE) Conference, Vol. 4836, Survey and Other Telescope Technologies and
  Discoveries. Edited by Tyson, J. Anthony; Wolff, Sidney. Proceedings of the
  SPIE, Volume 4836, pp. 61-72 (2002)., ed. J.~A. {Tyson} \& S.~{Wolff}, 61--72

\bibitem[{{Astier} {et~al.}(2006){Astier}, {Guy}, {Regnault}, {Pain},
  {Aubourg}, {Balam}, {Basa}, {Carlberg}, {Fabbro}, {Fouchez}, {Hook},
  {Howell}, {Lafoux}, {Neill}, {Palanque-Delabrouille}, {Perrett}, {Pritchet},
  {Rich}, {Sullivan}, {Taillet}, {Aldering}, {Antilogus}, {Arsenijevic},
  {Balland}, {Baumont}, {Bronder}, {Courtois}, {Ellis}, {Filiol}, {Gon{\c
  c}alves}, {Goobar}, {Guide}, {Hardin}, {Lusset}, {Lidman}, {McMahon},
  {Mouchet}, {Mourao}, {Perlmutter}, {Ripoche}, {Tao}, \&
  {Walton}}]{astier_etal06}
{Astier}, P., {Guy}, J., {Regnault}, N., {Pain}, R., {Aubourg}, E., {Balam},
  D., {Basa}, S., {Carlberg}, R.~G., {Fabbro}, S., {Fouchez}, D., {Hook},
  I.~M., {Howell}, D.~A., {Lafoux}, H., {Neill}, J.~D.,
  {Palanque-Delabrouille}, N., {Perrett}, K., {Pritchet}, C.~J., {Rich}, J.,
  {Sullivan}, M., {Taillet}, R., {Aldering}, G., {Antilogus}, P.,
  {Arsenijevic}, V., {Balland}, C., {Baumont}, S., {Bronder}, J., {Courtois},
  H., {Ellis}, R.~S., {Filiol}, M., {Gon{\c c}alves}, A.~C., {Goobar}, A.,
  {Guide}, D., {Hardin}, D., {Lusset}, V., {Lidman}, C., {McMahon}, R.,
  {Mouchet}, M., {Mourao}, A., {Perlmutter}, S., {Ripoche}, P., {Tao}, C., \&
  {Walton}, N. 2006, \aap, 447, 31

\bibitem[{{Barber}(2000)}]{barber00}
{Barber}, A.~J. 2000, \mnras, 318, 195

\bibitem[{{Bernardeau} {et~al.}(1997){Bernardeau}, {van Waerbeke}, \&
  {Mellier}}]{bernardeau_etal97}
{Bernardeau}, F., {van Waerbeke}, L., \& {Mellier}, Y. 1997, \aap, 322, 1

\bibitem[{{Christensen} \& {Meyer}(2001)}]{christensen_meyer01}
{Christensen}, N. \& {Meyer}, R. 2001, \prd, 64, 022001

\bibitem[{{Christensen} {et~al.}(2001){Christensen}, {Meyer}, {Knox}, \&
  {Luey}}]{christensen_etal01}
{Christensen}, N., {Meyer}, R., {Knox}, L., \& {Luey}, B. 2001, Classical and
  Quantum Gravity, 18, 2677

\bibitem[{{Cooray} {et~al.}(2006){Cooray}, {Holz}, \&
  {Huterer}}]{cooray_etal06}
{Cooray}, A., {Holz}, D.~E., \& {Huterer}, D. 2006, \apjl, 637, L77

\bibitem[{{Copin} {et~al.}(2006){Copin}, {Blanc}, {Bongard}, {Gangler},
  {Saug{\'e}}, {Smadja}, {Antilogus}, {Garavini}, {Gilles}, {Pain}, {Aldering},
  {Bailey}, {Lee}, {Loken}, {Nugent}, {Perlmutter}, {Scalzo}, {Thomas}, {Wang},
  {Weaver}, {P{\'e}contal}, {Kessler}, {Baltay}, {Rabinowitz}, \&
  {Bauer}}]{copin_etal06}
{Copin}, Y., {Blanc}, N., {Bongard}, S., {Gangler}, E., {Saug{\'e}}, L.,
  {Smadja}, G., {Antilogus}, P., {Garavini}, G., {Gilles}, S., {Pain}, R.,
  {Aldering}, G., {Bailey}, S., {Lee}, B.~C., {Loken}, S., {Nugent}, P.,
  {Perlmutter}, S., {Scalzo}, R., {Thomas}, R.~C., {Wang}, L., {Weaver}, B.~A.,
  {P{\'e}contal}, E., {Kessler}, R., {Baltay}, C., {Rabinowitz}, D., \&
  {Bauer}, A. 2006, New Astronomy Review, 50, 436

\bibitem[{{Dodelson} \& {Vallinotto}(2006)}]{dodelson_vallinotto06}
{Dodelson}, S. \& {Vallinotto}, A. 2006, \prd, 74, 063515

\bibitem[{{Guy} {et~al.}(2007){Guy}, {Astier}, {Baumont}, {Hardin}, {Pain},
  {Regnault}, {Basa}, {Carlberg}, {Conley}, {Fabbro}, {Fouchez}, {Hook},
  {Howell}, {Perrett}, {Pritchet}, {Rich}, {Sullivan}, {Antilogus}, {Aubourg},
  {Bazin}, {Bronder}, {Filiol}, {Palanque-Delabrouille}, {Ripoche}, \&
  {Ruhlmann-Kleider}}]{guy_etal07}
{Guy}, J., {Astier}, P., {Baumont}, S., {Hardin}, D., {Pain}, R., {Regnault},
  N., {Basa}, S., {Carlberg}, R.~G., {Conley}, A., {Fabbro}, S., {Fouchez}, D.,
  {Hook}, I.~M., {Howell}, D.~A., {Perrett}, K., {Pritchet}, C.~J., {Rich}, J.,
  {Sullivan}, M., {Antilogus}, P., {Aubourg}, E., {Bazin}, G., {Bronder}, J.,
  {Filiol}, M., {Palanque-Delabrouille}, N., {Ripoche}, P., \&
  {Ruhlmann-Kleider}, V. 2007, \aap, 466, 11

\bibitem[{{Hagan} {et~al.}(2005){Hagan}, {Ma}, \& {Kravtsov}}]{hagan_etal05}
{Hagan}, B., {Ma}, C.-P., \& {Kravtsov}, A.~V. 2005, \apj, 633, 537

\bibitem[{{Hamuy} {et~al.}(1995){Hamuy}, {Phillips}, {Maza}, {Suntzeff},
  {Schommer}, \& {Aviles}}]{hamuy_etal95}
{Hamuy}, M., {Phillips}, M.~M., {Maza}, J., {Suntzeff}, N.~B., {Schommer},
  R.~A., \& {Aviles}, R. 1995, \aj, 109, 1

\bibitem[{{Hannestad} {et~al.}(2008){Hannestad}, {Haugb{\o}lle}, \&
  {Thomsen}}]{hannestad_etal08}
{Hannestad}, S., {Haugb{\o}lle}, T., \& {Thomsen}, B. 2008, Journal of
  Cosmology and Astro-Particle Physics, 2, 22

\bibitem[{{Holz}(1998)}]{holz98}
{Holz}, D.~E. 1998, \apjl, 506, L1

\bibitem[{{Holz} \& {Linder}(2005)}]{holz_linder05}
{Holz}, D.~E. \& {Linder}, E.~V. 2005, \apj, 631, 678

\bibitem[{{Howell}(2001)}]{howell01}
{Howell}, D.~A. 2001, \apjl, 554, L193

\bibitem[{{Howell} {et~al.}(2007){Howell}, {Sullivan}, {Conley}, \&
  {Carlberg}}]{howell_etal07}
{Howell}, D.~A., {Sullivan}, M., {Conley}, A., \& {Carlberg}, R. 2007, \apjl,
  667, L37

\bibitem[{{Hu} \& {Jain}(2004)}]{hu_jain04}
{Hu}, W. \& {Jain}, B. 2004, \prd, 70, 043009

\bibitem[{{Huterer} {et~al.}(2004){Huterer}, {Kim}, {Krauss}, \&
  {Broderick}}]{huterer_etal04}
{Huterer}, D., {Kim}, A., {Krauss}, L.~M., \& {Broderick}, T. 2004, \apj, 615,
  595

\bibitem[{{Huterer} \& {Takada}(2005)}]{huterer_takada05}
{Huterer}, D. \& {Takada}, M. 2005, Astroparticle Physics, 23, 369

\bibitem[{{Huterer} \& {Turner}(2001)}]{huterer_turner01}
{Huterer}, D. \& {Turner}, M.~S. 2001, \prd, 64, 123527

\bibitem[{{Jha} {et~al.}(2007){Jha}, {Riess}, \& {Kirshner}}]{jha_etal07}
{Jha}, S., {Riess}, A.~G., \& {Kirshner}, R.~P. 2007, \apj, 659, 122

\bibitem[{{Jungman} {et~al.}(1996){Jungman}, {Kamionkowski}, {Kosowsky}, \&
  {Spergel}}]{jungman_etal96}
{Jungman}, G., {Kamionkowski}, M., {Kosowsky}, A., \& {Spergel}, D.~N. 1996,
  \prd, 54, 1332

\bibitem[{{Kantowski} {et~al.}(1995){Kantowski}, {Vaughan}, \&
  {Branch}}]{kantowski_etal95}
{Kantowski}, R., {Vaughan}, T., \& {Branch}, D. 1995, \apj, 447, 35

\bibitem[{{Kim} {et~al.}(2004){Kim}, {Linder}, {Miquel}, \&
  {Mostek}}]{kim_etal04}
{Kim}, A.~G., {Linder}, E.~V., {Miquel}, R., \& {Mostek}, N. 2004, \mnras, 347,
  909

\bibitem[{{Knox} {et~al.}(2001){Knox}, {Christensen}, \&
  {Skordis}}]{knox_etal01}
{Knox}, L., {Christensen}, N., \& {Skordis}, C. 2001, \apjl, 563, L95

\bibitem[{{Komatsu} {et~al.}(2008){Komatsu}, {Dunkley}, {Nolta}, {Bennett},
  {Gold}, {Hinshaw}, {Jarosik}, {Larson}, {Limon}, {Page}, {Spergel},
  {Halpern}, {Hill}, {Kogut}, {Meyer}, {Tucker}, {Weiland}, {Wollack}, \&
  {Wright}}]{komatsu_etal08}
{Komatsu}, E., {Dunkley}, J., {Nolta}, M.~R., {Bennett}, C.~L., {Gold}, B.,
  {Hinshaw}, G., {Jarosik}, N., {Larson}, D., {Limon}, M., {Page}, L.,
  {Spergel}, D.~N., {Halpern}, M., {Hill}, R.~S., {Kogut}, A., {Meyer}, S.~S.,
  {Tucker}, G.~S., {Weiland}, J.~L., {Wollack}, E., \& {Wright}, E.~L. 2008,
  ApJS Submitted, ArXiv e-prints, 803

\bibitem[{{Kosowsky} {et~al.}(2002){Kosowsky}, {Milosavljevic}, \&
  {Jimenez}}]{kosowsky_etal02}
{Kosowsky}, A., {Milosavljevic}, M., \& {Jimenez}, R. 2002, \prd, 66, 063007

\bibitem[{{Linder} {et~al.}(1988){Linder}, {Wagoner}, \&
  {Schneider}}]{linder_etal88}
{Linder}, E.~V., {Wagoner}, R.~V., \& {Schneider}, P. 1988, \apj, 324, 786

\bibitem[{{Ma} {et~al.}(2006){Ma}, {Hu}, \& {Huterer}}]{ma_etal06}
{Ma}, Z., {Hu}, W., \& {Huterer}, D. 2006, \apj, 636, 21

\bibitem[{{Mannucci} {et~al.}(2006){Mannucci}, {Della Valle}, \&
  {Panagia}}]{mannucci_etal06}
{Mannucci}, F., {Della Valle}, M., \& {Panagia}, N. 2006, \mnras, 370, 773

\bibitem[{{Metcalf}(1999)}]{metcalf99}
{Metcalf}, R.~B. 1999, \mnras, 305, 746

\bibitem[{{Munshi} \& {Jain}(2000)}]{munshi_jain00}
{Munshi}, D. \& {Jain}, B. 2000, \mnras, 318, 109

\bibitem[{{Phillips}(1993)}]{phillips93}
{Phillips}, M.~M. 1993, \apjl, 413, L105

\bibitem[{{Pinto} {et~al.}(2004){Pinto}, {Smith}, \&
  {Garnavich}}]{pinto_etal04}
{Pinto}, P.~A., {Smith}, C.~R., \& {Garnavich}, P.~M. 2004, BAAS, 36, 1530

\bibitem[{{Prieto} {et~al.}(2006){Prieto}, {Rest}, \&
  {Suntzeff}}]{prieto_etal06}
{Prieto}, J.~L., {Rest}, A., \& {Suntzeff}, N.~B. 2006, \apj, 647, 501

\bibitem[{{Riess} {et~al.}(1996){Riess}, {Press}, \& {Kirshner}}]{riess_etal96}
{Riess}, A.~G., {Press}, W.~H., \& {Kirshner}, R.~P. 1996, \apj, 473, 88

\bibitem[{{Riess} {et~al.}(2007){Riess}, {Strolger}, {Casertano}, {Ferguson},
  {Mobasher}, {Gold}, {Challis}, {Filippenko}, {Jha}, {Li}, {Tonry}, {Foley},
  {Kirshner}, {Dickinson}, {MacDonald}, {Eisenstein}, {Livio}, {Younger}, {Xu},
  {Dahl{\'e}n}, \& {Stern}}]{riess_etal07}
{Riess}, A.~G., {Strolger}, L.-G., {Casertano}, S., {Ferguson}, H.~C.,
  {Mobasher}, B., {Gold}, B., {Challis}, P.~J., {Filippenko}, A.~V., {Jha}, S.,
  {Li}, W., {Tonry}, J., {Foley}, R., {Kirshner}, R.~P., {Dickinson}, M.,
  {MacDonald}, E., {Eisenstein}, D., {Livio}, M., {Younger}, J., {Xu}, C.,
  {Dahl{\'e}n}, T., \& {Stern}, D. 2007, \apj, 659, 98

\bibitem[{{Rudd} {et~al.}(2008){Rudd}, {Zentner}, \& {Kravtsov}}]{rudd_etal08}
{Rudd}, D.~H., {Zentner}, A.~R., \& {Kravtsov}, A.~V. 2008, \apj, 672, 19

\bibitem[{{Sarkar} {et~al.}(2008{\natexlab{a}}){Sarkar}, {Amblard}, {Cooray},
  \& {Holz}}]{sarkar_etal08b}
{Sarkar}, D., {Amblard}, A., {Cooray}, A., \& {Holz}, D.~E. 2008{\natexlab{a}},
  ArXiv e-prints, 806

\bibitem[{{Sarkar} {et~al.}(2008{\natexlab{b}}){Sarkar}, {Amblard}, {Holz}, \&
  {Cooray}}]{sarkar_etal08a}
{Sarkar}, D., {Amblard}, A., {Holz}, D.~E., \& {Cooray}, A. 2008{\natexlab{b}},
  \apj, 678, 1

\bibitem[{{Sasaki}(1987)}]{sasaki87}
{Sasaki}, M. 1987, \mnras, 228, 653

\bibitem[{{Schneider} \& {Wagoner}(1987)}]{schneider_wagoner87}
{Schneider}, P. \& {Wagoner}, R.~V. 1987, \apj, 314, 154

\bibitem[{{Seljak}(1997)}]{seljak97}
{Seljak}, U. 1997, \apj, 482, 6

\bibitem[{{Smith} {et~al.}(2003){Smith}, {Peacock}, {Jenkins}, {White},
  {Frenk}, {Pearce}, {Thomas}, {Efstathiou}, \& {Couchman}}]{smith_etal03}
{Smith}, R.~E., {Peacock}, J.~A., {Jenkins}, A., {White}, S.~D.~M., {Frenk},
  C.~S., {Pearce}, F.~R., {Thomas}, P.~A., {Efstathiou}, G., \& {Couchman},
  H.~M.~P. 2003, \mnras, 341, 1311

\bibitem[{{Sullivan} {et~al.}(2006){Sullivan}, {Le Borgne}, {Pritchet},
  {Hodsman}, {Neill}, {Howell}, {Carlberg}, {Astier}, {Aubourg}, {Balam},
  {Basa}, {Conley}, {Fabbro}, {Fouchez}, {Guy}, {Hook}, {Pain},
  {Palanque-Delabrouille}, {Perrett}, {Regnault}, {Rich}, {Taillet}, {Baumont},
  {Bronder}, {Ellis}, {Filiol}, {Lusset}, {Perlmutter}, {Ripoche}, \&
  {Tao}}]{sullivan_etal06}
{Sullivan}, M., {Le Borgne}, D., {Pritchet}, C.~J., {Hodsman}, A., {Neill},
  J.~D., {Howell}, D.~A., {Carlberg}, R.~G., {Astier}, P., {Aubourg}, E.,
  {Balam}, D., {Basa}, S., {Conley}, A., {Fabbro}, S., {Fouchez}, D., {Guy},
  J., {Hook}, I., {Pain}, R., {Palanque-Delabrouille}, N., {Perrett}, K.,
  {Regnault}, N., {Rich}, J., {Taillet}, R., {Baumont}, S., {Bronder}, J.,
  {Ellis}, R.~S., {Filiol}, M., {Lusset}, V., {Perlmutter}, S., {Ripoche}, P.,
  \& {Tao}, C. 2006, \apj, 648, 868

\bibitem[{{Tegmark} {et~al.}(1997){Tegmark}, {Taylor}, \&
  {Heavens}}]{tegmark_etal97}
{Tegmark}, M., {Taylor}, A.~N., \& {Heavens}, A.~F. 1997, \apj, 480, 22

\bibitem[{{Valageas}(2000)}]{valageas00}
{Valageas}, P. 2000, \aap, 356, 771

\bibitem[{{Wambsganss} {et~al.}(1997){Wambsganss}, {Cen}, {Xu}, \&
  {Ostriker}}]{wambsganss_etal97}
{Wambsganss}, J., {Cen}, R., {Xu}, G., \& {Ostriker}, J.~P. 1997, \apjl, 475,
  L81+

\bibitem[{{Wang}(1999)}]{wang99}
{Wang}, Y. 1999, \apj, 525, 651

\bibitem[{{Wang} {et~al.}(2002){Wang}, {Holz}, \& {Munshi}}]{wang_etal02}
{Wang}, Y., {Holz}, D.~E., \& {Munshi}, D. 2002, \apjl, 572, L15

\bibitem[{{Wang} {et~al.}(2007){Wang}, {Narayan}, \&
  {Wood-Vasey}}]{wang_etal07}
{Wang}, Y., {Narayan}, G., \& {Wood-Vasey}, M. 2007, MNRAS, 382, 377

\bibitem[{{White}(2004)}]{white04}
{White}, M. 2004, Astroparticle Physics, 22, 211

\bibitem[{{Wood-Vasey} {et~al.}(2007){Wood-Vasey}, {Miknaitis}, {Stubbs},
  {Jha}, {Riess}, {Garnavich}, {Kirshner}, {Aguilera}, {Becker}, {Blackman},
  {Blondin}, {Challis}, {Clocchiatti}, {Conley}, {Covarrubias}, {Davis},
  {Filippenko}, {Foley}, {Garg}, {Hicken}, {Krisciunas}, {Leibundgut}, {Li},
  {Matheson}, {Miceli}, {Narayan}, {Pignata}, {Prieto}, {Rest}, {Salvo},
  {Schmidt}, {Smith}, {Sollerman}, {Spyromilio}, {Tonry}, {Suntzeff}, \&
  {Zenteno}}]{wood-vasey_etal07}
{Wood-Vasey}, W.~M., {Miknaitis}, G., {Stubbs}, C.~W., {Jha}, S., {Riess},
  A.~G., {Garnavich}, P.~M., {Kirshner}, R.~P., {Aguilera}, C., {Becker},
  A.~C., {Blackman}, J.~W., {Blondin}, S., {Challis}, P., {Clocchiatti}, A.,
  {Conley}, A., {Covarrubias}, R., {Davis}, T.~M., {Filippenko}, A.~V.,
  {Foley}, R.~J., {Garg}, A., {Hicken}, M., {Krisciunas}, K., {Leibundgut}, B.,
  {Li}, W., {Matheson}, T., {Miceli}, A., {Narayan}, G., {Pignata}, G.,
  {Prieto}, J.~L., {Rest}, A., {Salvo}, M.~E., {Schmidt}, B.~P., {Smith},
  R.~C., {Sollerman}, J., {Spyromilio}, J., {Tonry}, J.~L., {Suntzeff}, N.~B.,
  \& {Zenteno}, A. 2007, \apj, 666, 694

\bibitem[{{Zentner} {et~al.}(2008){Zentner}, {Rudd}, \& {Hu}}]{zentner_etal08}
{Zentner}, A.~R., {Rudd}, D.~H., \& {Hu}, W. 2008, \prd, 77, 043507

\bibitem[{{Zhan} \& {Knox}(2004)}]{zhan_knox04}
{Zhan}, H. \& {Knox}, L. 2004, \apjl, 616, L75

\bibitem[{{Zhan} {et~al.}(2008){Zhan}, {Wang}, {Pinto}, \&
  {Tyson}}]{zhan_etal08}
{Zhan}, H., {Wang}, L., {Pinto}, P., \& {Tyson}, J.~A. 2008, \apjl, 675, L1

\bibitem[{{Zhang} \& {Chen}(2008)}]{zhang_chen08}
{Zhang}, P. \& {Chen}, X. 2008, \prd, 78, 023006

\end{thebibliography}

\end{document}